\documentclass[iop]{emulateapj}
\usepackage{amsmath, amsfonts,amssymb,natbib,graphicx,color,subfigure}

\begin{document}
\title{Mapping dust through emission and absorption in nearby galaxies}

\author{Kathryn Kreckel\altaffilmark{1},  Brent Groves\altaffilmark{1}, Eva Schinnerer\altaffilmark{1}, Benjamin D. Johnson\altaffilmark{2}, Gonzalo Aniano\altaffilmark{3}, Daniela Calzetti\altaffilmark{4}, Kevin V. Croxall\altaffilmark{5}, Bruce T. Draine\altaffilmark{6}, Karl D. Gordon\altaffilmark{7}, Alison F. Crocker\altaffilmark{8}, Daniel A. Dale\altaffilmark{9}, Leslie K. Hunt\altaffilmark{10}, Robert C. Kennicutt\altaffilmark{11}, Sharon E. Meidt\altaffilmark{1}, J. D. T. Smith\altaffilmark{8}, Fatemeh S. Tabatabaei\altaffilmark{1}}

\altaffiltext{1}{Max Planck Institut f\"{u}r Astronomie, K\"{o}nigstuhl 17, 69117 Heidelberg, Germany;  kreckel@mpia.de}
\altaffiltext{2}{Institut d'Astrophysique de Paris, UMR 7095, 98 bis Bvd Arago, 75014 Paris, France}
\altaffiltext{3}{Institut d'Astrophysique Spatiale (IAS), b\^{a}timent 121, Universit\'{e} Paris-Sud 11 and CNRS (UMR 8617), F-91405 Orsay, France}
\altaffiltext{4}{Department of Astronomy, University of Massachusetts, Amherst, MA 01003, USA}
\altaffiltext{5}{Department of Astronomy, The Ohio State University, 140 West 18th Avenue, Columbus, OH 43210, USA}
\altaffiltext{6}{Princeton University Observatory, Peyton Hall, Princeton, NJ 08544-1001, USA}
\altaffiltext{7}{Space Telescope Science Institute, 3700 San Martin Drive, Baltimore, MD 21218, USA}
\altaffiltext{8}{Department of Physics and Astronomy, University of Toledo, Toledo, OH 43606, USA}
\altaffiltext{9}{Department of Physics and Astronomy, University of Wyoming, Laramie, WY 82071, USA}
\altaffiltext{10}{INAF-Osservatorio Astroﬁsico di Arcetri, Largo E. Fermi 5, I-50125 Firenze, Italy}
\altaffiltext{11}{Institute of Astronomy, University of Cambridge, Madingley Road, Cambridge CB3 0HA, UK}

\begin{abstract}
Dust has long been identified as a barrier to measuring inherent galaxy properties. However, the link between dust and attenuation is not straightforward and depends on both the amount of dust and its distribution. 
 Herschel imaging of nearby galaxies undertaken as part of the KINGFISH project allows us to map the dust as seen in emission with unprecedented sensitivity and $\sim$1 kpc resolution. 
We present here new optical integral field unit spectroscopy for eight of these galaxies that provides complementary 100-200 pc scale maps of the dust attenuation through observation of the reddening in both the Balmer decrement and the stellar continuum. The stellar continuum reddening, which is systematically less than that observed in the Balmer decrement, shows no clear correlation with the dust, suggesting that the distribution of stellar reddening acts as a poor tracer of the overall dust content. 
The brightest H\textsc{ii} regions are observed to be preferentially located in dusty regions, and we do find a correlation between the Balmer line reddening and the dust mass surface density for which we provide an empirical relation.   Some of the high-inclination systems in our sample exhibit high extinction, but we also find evidence that unresolved variations in the dust distribution on scales smaller than 500 pc may contribute to the scatter in this relation. We caution against the use of integrated A$_V$ measures to infer global dust properties.
\end{abstract}

\section{Introduction} 
The production of interstellar dust is closely tied to the formation and destruction of stars \citep{Dwek1998,Franceschini2000}, and is expected to be present even in the early universe after being created by the ejecta of population III stars \citep{Nozawa2003}.  Through the absorption and scattering of optical and UV photons, dust affects both the intrinsic continuum emission from stars and the line and continuum emission from nebulae, resulting in both extinction (A$_V$) and reddening (E(B-V)) of the light.  In studies of galaxies it can significantly impact derived quantities such as star formation rate (SFR), mean stellar age and mass-to-light ratio, and therefore must be taken into account. 

Corrections for these effects at optical wavelengths are typically done assuming an extinction law \citep{Cardelli1989, ODonnell1994, Fitzpatrick1999}, appropriate for absorption and scattering out of the line of sight by dust , or an attenuation law \citep{Calzetti2000}, which applies to mixed dust-source geometries that include scattering both out of and into the line of sight.  These laws are parameterized by the ratio of total to selective extinction, R$_V$ = A$_V$/E(B-V), where the value of R$_V$ depends upon the ISM conditions where extinction is occurring but is typically 3.1 in the Milky Way.  While the connection between dust and extinction is relatively simple for a single illuminating source, the picture becomes more complex when we consider multiple unresolved sources within a galaxy.  Sources of different luminosities may probe different optical depths within a single region, and will selectively sample the dust distribution depending on their locations.  One example of this effect is in H\textsc{ii} regions, which are expected to be preferentially more buried than the typical stellar population due to the association of the ionizing O stars with the gas from which they formed \citep{Calzetti1994, Calzetti2000, Charlot2000}. This association biases V-band extinction measures (A$_V$) based on gas emission line reddening (A$_{V,g}$) to higher values than the extinction determined from the reddening of the stellar continuum light (A$_{V,s}$). 

However, dust is also visible through its thermal emission in the infrared, and with the advent of space based infrared observatories such as the Infrared Astronomical Satellite (IRAS) and the Infrared Space Observatory (ISO) it became possible to measure the dust directly, without the limitation of geometric effects.  Using either blackbody fits \citep[e.g.][]{Galametz2012} or more detailed SED fitting assuming a specific dust composition \citep[e.g.][]{Aniano2012}, it is possible to map the total dust distribution within galaxies. Initially possible at high spatial resolution for only the closest galaxies (the Magellanic Clouds, M31, M33),  the latest generation of infrared telescopes has enabled high spatial resolution studies of dust emission in a representative sample of nearby galaxies. 

In this work we compare the total amount of dust, as seen in emission, with the optical attenuation by dust, as seen by reddening of both the stellar continuum and the Hydrogen Balmer emission lines.  We base our analysis on a subsample of galaxies taken from the Key Insights on Nearby Galaxies: A Far-Infrared Survey with Herschel (KINGFISH, \citealt{Kennicutt2011}) project.  Given the 18\arcsec\ angular resolution achieved by Herschel at 250$\,\mu$m, the large angular diameter of these targets allows us to distinguish features at physical scales of $\sim$1 kpc within the galaxy disks.  We complement these data with optical integral field spectrograph (IFS) observations that provide complete sampling of a $\sim$1\arcmin~ field of view by 2\farcs7 fibers.  We derive an empirical relation between the total dust column and optical extinction, and through simple models we explore the effects of geometry and quantify biases in the optical tracers.

We present an overview of the data in Section \ref{sec:data}.  In Section \ref{sec:results} we present results of our comparison of the Balmer decrement with the dust mass inferred from the far-IR emission, as well as a comparison of the optical reddening of the stellar continuum with the Balmer emission lines.  Finally, we discuss implications for the dust geometry in Section \ref{sec:discussion}, and provide an empirical relation between dust absorption and emission. We conclude in Section \ref{sec:conclusion}.


\begin{deluxetable*}{l c c c c c c c c c}[ht!]
\tabletypesize{\small}
\tablecaption{Galaxy parameters taken from \cite{Kennicutt2011}. 
\label{tab:properties}}
\tablewidth{0pt}
\tablehead{
\colhead{Name} & \colhead{Morphology} & \colhead{Distance} & \colhead{Nuclear type} & \colhead{SFR} & \colhead{Log[M$_*$]} & \colhead{12+log(O/H)} & \colhead{12+log(O/H)} & \colhead{inclination} & \colhead{A$_V$ (MW)} \\
\colhead{} & \colhead{} & \colhead{(Mpc)} & \colhead{} & \colhead{(M$_\sun$ yr$^{-2}$)} & \colhead{(M$_\sun$)} & \colhead{[PT05]\tablenotemark{a}} & \colhead{[KK04]\tablenotemark{b}} & \colhead{($^\circ$)} & \colhead{(mag)}
} 
\startdata
NGC 2146 & SBab &  17.2	& SF & 7.94 & 10.30 & 8.68 & ... & 60  \tablenotemark{c} & 0.264 \\
NGC 2798 & SBa &  25.8 & SF/AGN & 3.38 & 10.04 &  8.34 &	9.04 & 68 \tablenotemark{d,*} & 0.055 \\
NGC 3077 & I0pec &  3.83 &	SF & 0.094 & 9.34 & ... & 8.9 &  46 \tablenotemark{e} & 0.184\\
NGC 3627 & SABb &   9.38 & 	AGN  & 1.70  & 10.49 & 	8.34 & 	8.99 &  62 \tablenotemark{e} & 0.089\\
NGC 4321 & 	SABbc &   14.3 &  	AGN  &  2.61 & 10.30 &  	8.50 &  	9.17 &  32  \tablenotemark{d} & 0.072 \\ 

NGC 5713 & SABbcp  &   21.4 &    SF  &  2.52  & 10.07 &  8.24 &  	9.03 &  33  \tablenotemark{d} & 0.108 \\ 
NGC 6946 & SABcd  &  	6.8  &   SF  &   7.12  & 9.96 &  	8.40  &  	9.05 &  33 \tablenotemark{e}  & 0.937 \\
NGC 7331 & SAb  & 14.5  & AGN  & 2.74  & 10.56 & 8.34 & 	9.02 & 76 \tablenotemark{e}  & 0.249 \\
\enddata
\tablenotetext{a}{\cite{PT05}}
\tablenotetext{b}{\cite{KK04}}
\tablenotetext{c}{\cite{Garrido2005}}
\tablenotetext{d}{\cite{Daigle2006}}
\tablenotetext{e}{\cite{Walter2008}}
\tablenotetext{*}{Photometrically determined}
\end{deluxetable*}

\section{Data} 
\label{sec:data}

\subsection{Sample Selection} 
For this work we have selected eight galaxies from the KINGFISH survey of 61 nearby galaxies.  Our main criterion was that these sources be bright in CO, with the future goal of studying excitation in the multi-phase ISM in conjunction with a SPIRE FTS program (PI: J.D. Smith), and some care was taken to ensure they span a moderate range of Hubble type (irregular and spiral) and nuclear excitation sources (starburst, LINER and AGN).   General properties of these galaxies are given in Table \ref{tab:properties}. We list only inclinations measured using 2D gas kinematic information (H$\alpha$ or HI velocity fields), except for NGC 2798 where none was available.  We also show the A$_V$ of the foreground Milky Way extinction  that we assume for each galaxy position based on the \cite{Cardelli1989} model using the \cite{Schlafly2011} revisions to the \cite{Schlegel1998} dust maps.  Only NGC 6946 falls in a region of the sky with large foreground extinction due to the Milky Way, which we correct for in our spectral data cube (see Section \ref{sec:linemaps}).

Figure \ref{fig:optical} presents optical R-band images of the galaxies and indicates our targeted IFS field of view.  NGC 2146 and NGC 7331 have particularly strong dust lane features.  About half of our selected galaxies are at high inclination, above 60$^\circ$, while the rest are closer to face-on.    NGC 2798, 3627, 4321 and 7331 exhibit evidence of a central AGN, although none is bright enough to dominate the stellar continuum light in the galaxy.  NGC 2146 is classified as a luminous infrared galaxy (LIRG) \citep{Sanders2003} and has been identified as having strong stellar winds as part of a starburst, similar to M82.  
\\

\subsection{Far-IR: dust in emission} 
Using imaging from the PACS \citep{Poglitsch2010} and SPIRE \citep{Griffin2010} cameras on the Herschel Space Observatory \citep{Pilbratt2010} and the IRAC \citep{Fazio2004} and MIPS \citep{Rieke2004} cameras on the Spitzer Space Telescope \citep{Werner2004}, the dust SED in these galaxies has been physically modeled using the \cite{Draine2007} dust model to construct maps of the dust mass surface density, $\Sigma$M$_d$. This method as applied to two test galaxies, NGC 628 and NGC 6946, is presented in detail in \cite{Aniano2012}, and similar modeling for the complete KINGFISH sample will be presented in a future work.  The model specifies the dust properties, including a distribution of grain sizes, compositions and a frequency-dependent opacity, and fits the fraction of dust in polycyclic aromatic hydrocarbons, the starlight intensity which heats the dust and the mass surface density of dust within  resolved pixels at 200-700 pc scales.  Uncertainties in the dust mass are calculated taking into account both uncertainties in the imaging and in the model fitting. 

To achieve a reasonably high angular resolution, particularly considering the limited $\sim$1\arcmin~ field of view from our IFS observations (described below), we restrict ourselves to wavelengths up to and including 250 $\mu$m for these dust models to achieve a FWHM of 18\farcs2. Before fitting, images at all wavelengths have been convolved to the 250 $\mu$m resolution as described in \cite{Aniano2011}.  Models are fitted to data from the IRAC 3.6, 4.5, 5.8 and 8.0 $\mu$m images, the MIPS 24 $\mu$m image, the PACS 70, 100, and 160 $\mu$m images, and the SPIRE 250 $\mu$m image.  Detailed comparison of the resulting dust mass estimates  omitting the longest wavelength SED points suggest that this may result in an overestimate of the dust mass by approximately 35\% \citep{Aniano2012}.

\subsection{Optical: dust in absorption} 
\label{sec:data_optical}

\subsubsection{Observations} 

\begin{deluxetable}{l c c c c }[t]
\tabletypesize{\small}
\tablecaption{PPAK science targets observing details. 
\label{tab:ppak}}
\tablewidth{0pt}
\tablehead{\colhead{NGC} & \colhead{RA} & \colhead{Dec} & \colhead{Exp. time} & \colhead{Date} \\
\colhead{} & \colhead{} & \colhead{} & \colhead{(s)} & \colhead{} 
}
\startdata
 2146 & 6:18:37.3 & +78:21:33.0 & 3 $\times$ (3 $\times$ 420) & Feb 27, 2012 \\
 2798 & 9:17:22.9 & +42:00:05.1  & 3 $\times$ (3 $\times$ 420) & Feb 28, 2012 \\
 3077 &  10:03:17.8 & +68:44:12.9 & 3 $\times$ (3 $\times$ 420) & Feb 28, 2012 \\
 3627 & 11:20:12.0  & 12:59:07.4 & 3 $\times$ (3 $\times$ 420) & Feb 27, 2012 \\
 &  11:20:13.8 & +13:00:03.3 & 3 $\times$ (3 $\times$ 420) & Feb 28, 2012 \\
 &  11:20:15.5 & +12:59:02.1 & 3 $\times$ (3 $\times$ 420) & Feb 27, 2012 \\
 &  11:20:17.8 & +12:59:51.6 & 3 $\times$ (3 $\times$ 420) & Feb 27, 2012 \\
 4321 & 12:22:51.2 & +15:49:55.5  &  3 $\times$ (3 $\times$ 420) & Feb 28, 2012 \\
 &  12:22:54.7 & +15:46:24.5 & 3 $\times$ (3 $\times$ 420) & Feb 28, 2012 \\
  &  12:22:59.1 & +15:48:54.4 & 3 $\times$ (3 $\times$ 420) & Feb 28, 2012 \\
 5713 &  14:40:11.4 & -00:17:17.9 & 3 $\times$ (1200) & July 26, 2011 \\
 6946 & 20:34:52.4 & +60:09:17.0 & 3 $\times$ (1200) & July 26, 2011 \\
(6946E) & 20:35:12.0 & +60:08:59.5 & 3 $\times$ (1200) & July 26, 2011  \\
 7331 &  22:37:04.0 & +34:25:02.2 & 3 $\times$ (1200) & July 26, 2011  
\enddata
\end{deluxetable}

Optical IFU spectroscopy was obtained using the PMAS instrument \citep{Roth2005} in PPAK mode on the Calar Alto 3.5m telescope \citep{Kelz2006}.  Having 331 fibers each 2\farcs68 in diameter arranged in a hexagonal pattern, it provides a  $\sim$1\arcmin~ field of view. Gaps between the fibers result in a filling factor of about 60\%. Due to the large angular extent of our targets, additional fibers intended to simultaneously sample the sky are generally contaminated by galaxy emission, and so separate sky pointings were obtained.  
Details of the observations for all science targets are given in Table \ref{tab:ppak}.  We give positions for each pointing, including those mosaicked across the spiral and bar regions of NGC 3627 and NGC 4321.  Exposure times listed account for the three dither positions observed, as well as observations that were split into three exposures for cosmic ray removal.  Conditions were photometric, with typical sub-fiber seeing of less than 1\arcsec.  

\begin{figure*}[h]
\centering
\includegraphics[height=1.8in]{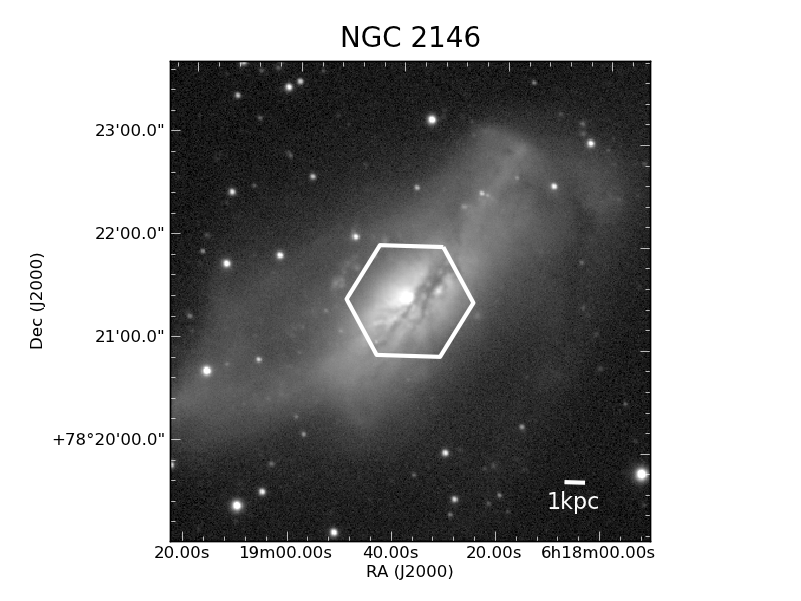}
\hspace{-.5cm}\includegraphics[height=1.6in]{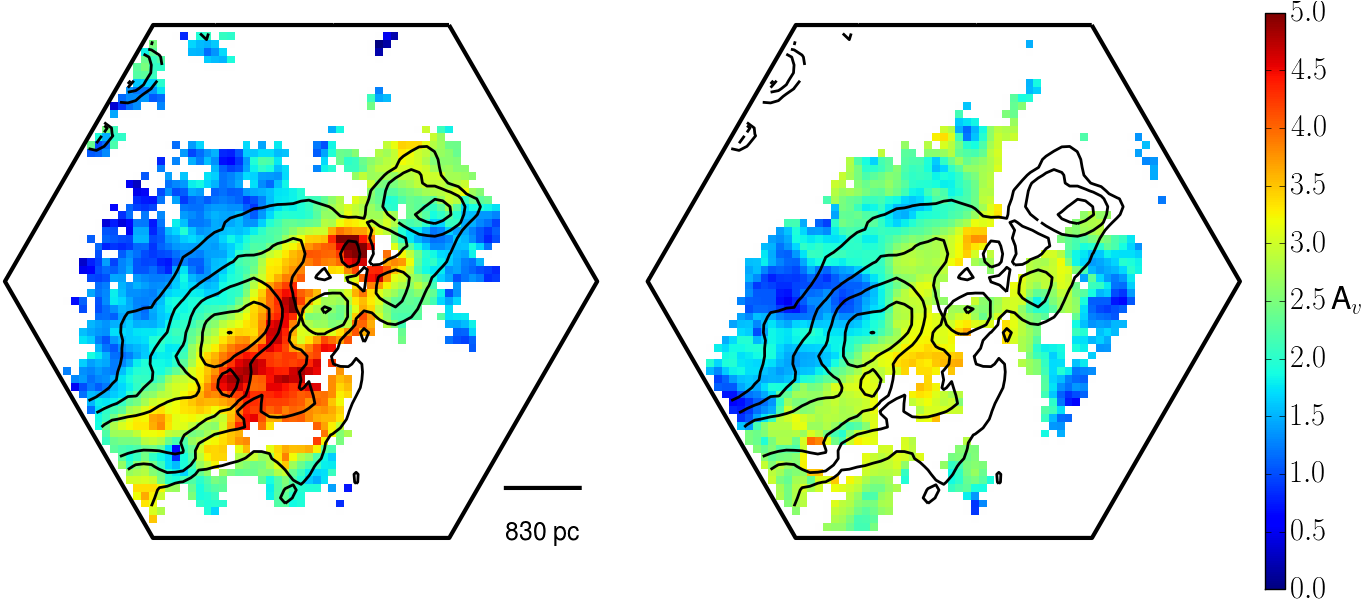}
\includegraphics[height=1.8in]{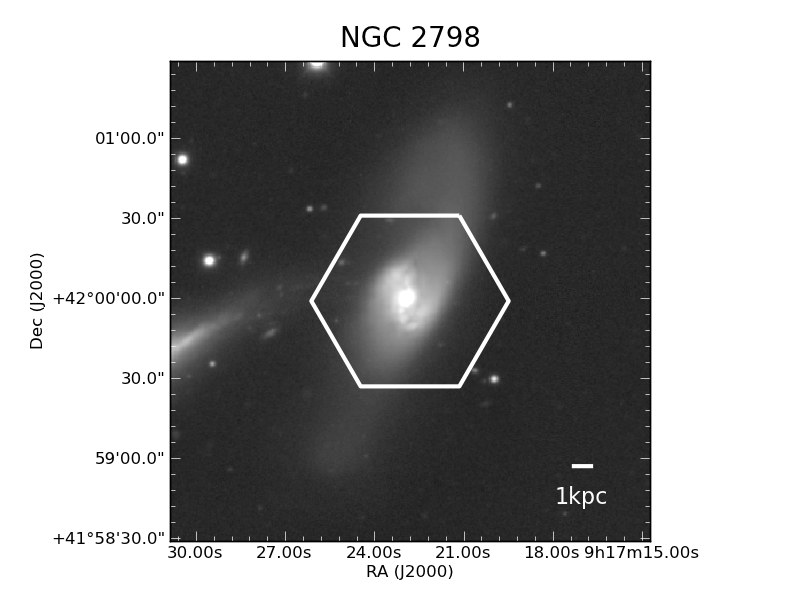}
\hspace{-.5cm}\includegraphics[height=1.6in]{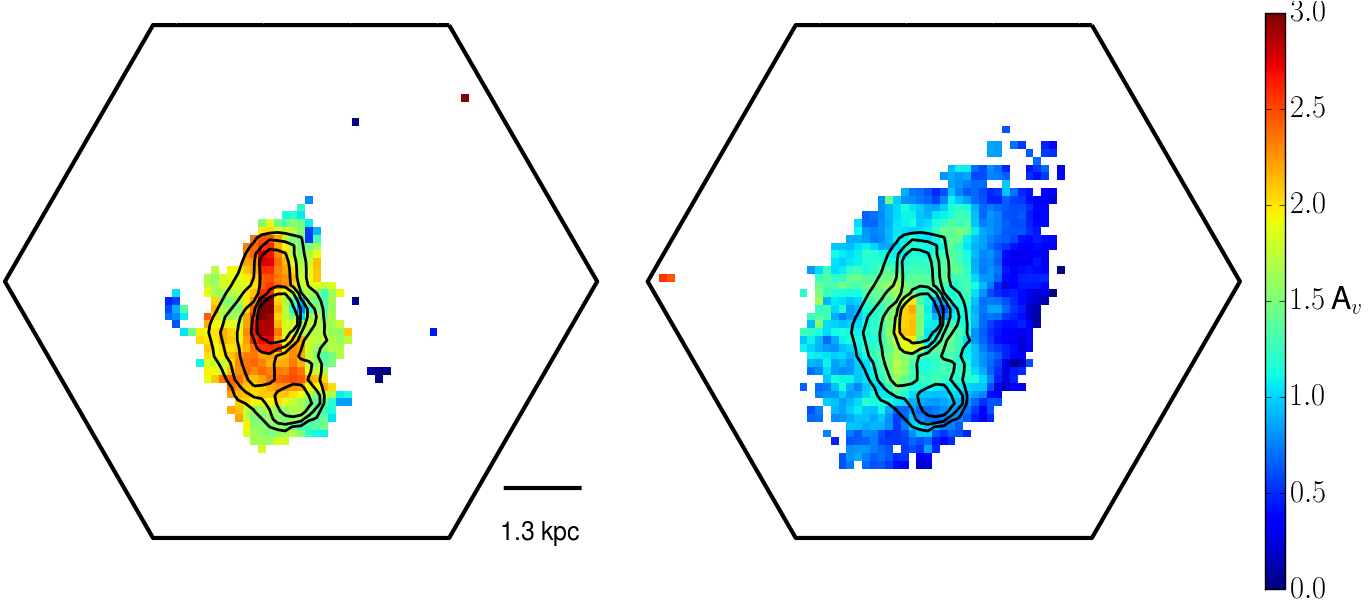}
\includegraphics[height=1.8in]{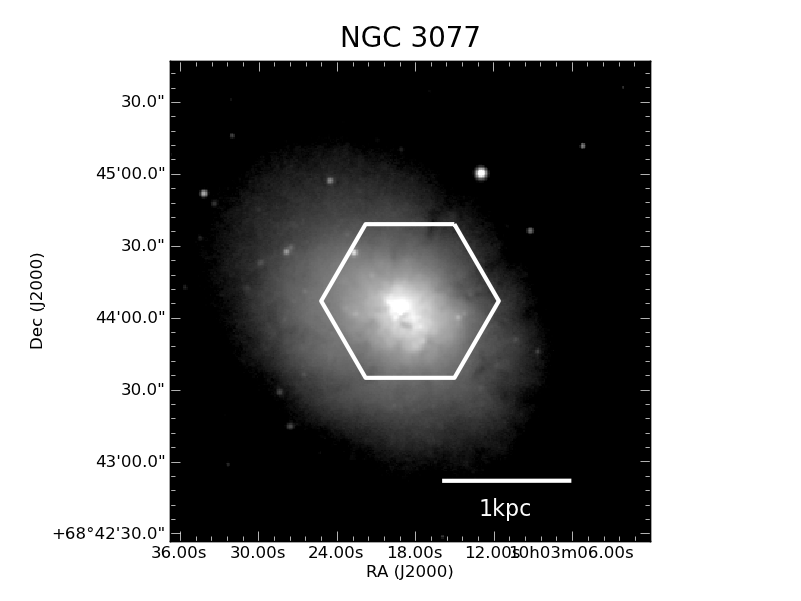}
\hspace{-.5cm}\includegraphics[height=1.6in]{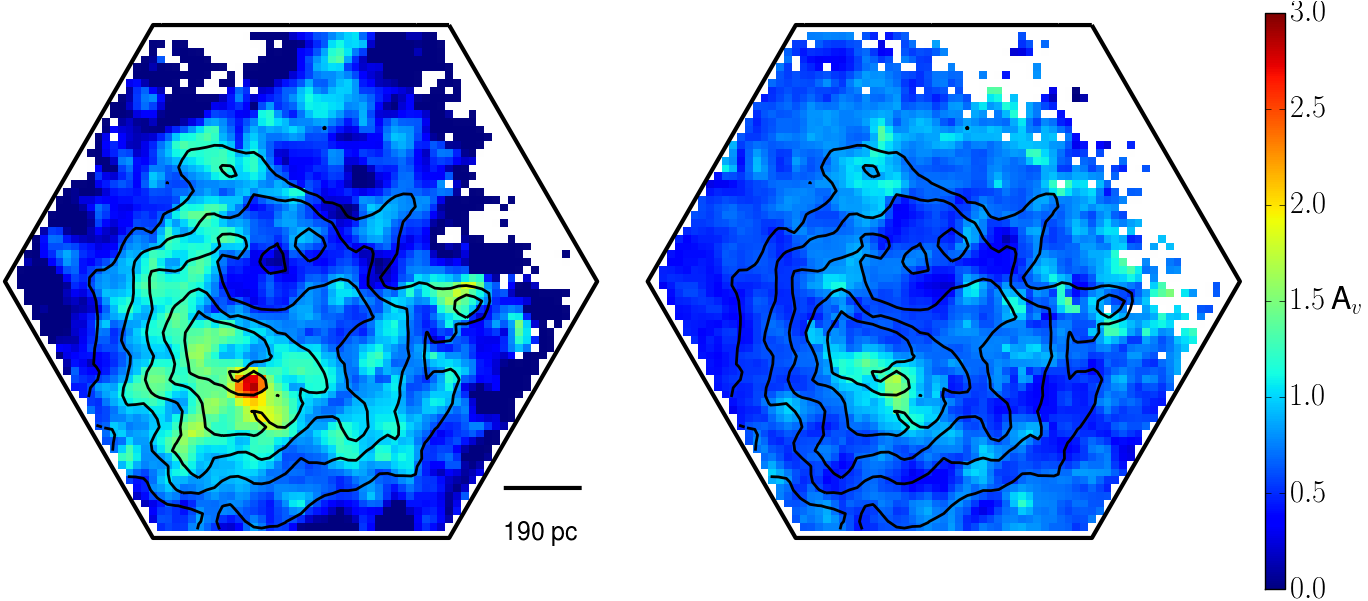}
\includegraphics[height=1.8in]{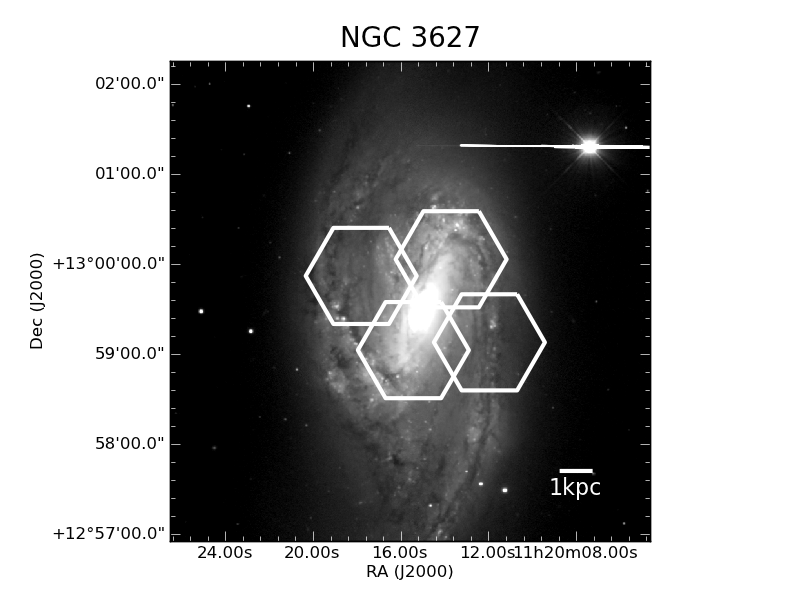}
\hspace{-.5cm}\includegraphics[height=1.6in]{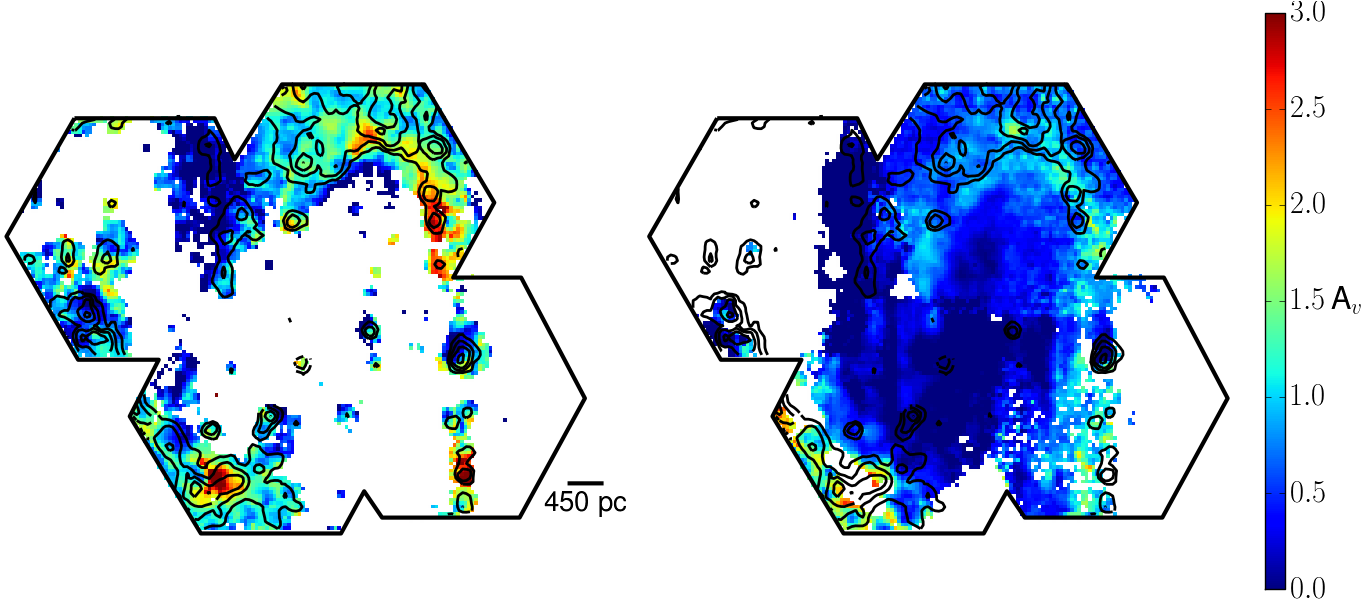}
\includegraphics[height=1.8in]{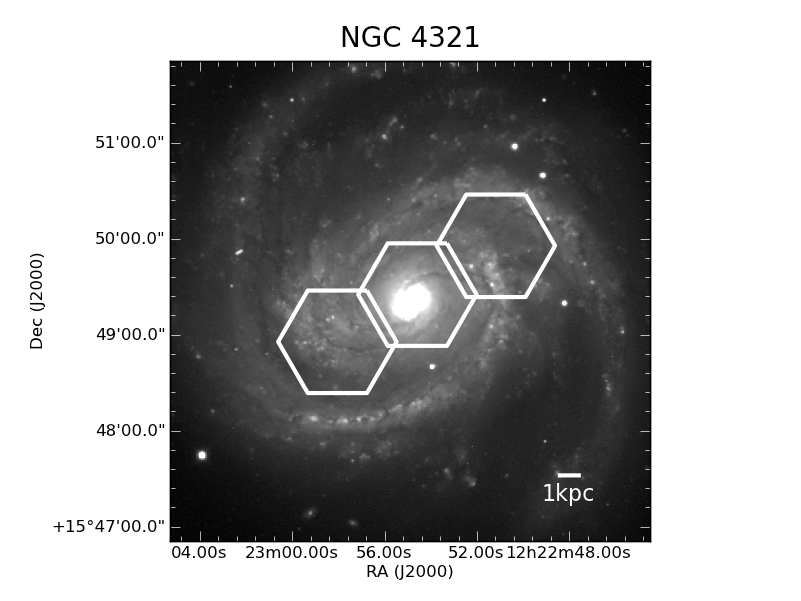}
\hspace{-.5cm}\includegraphics[height=1.6in]{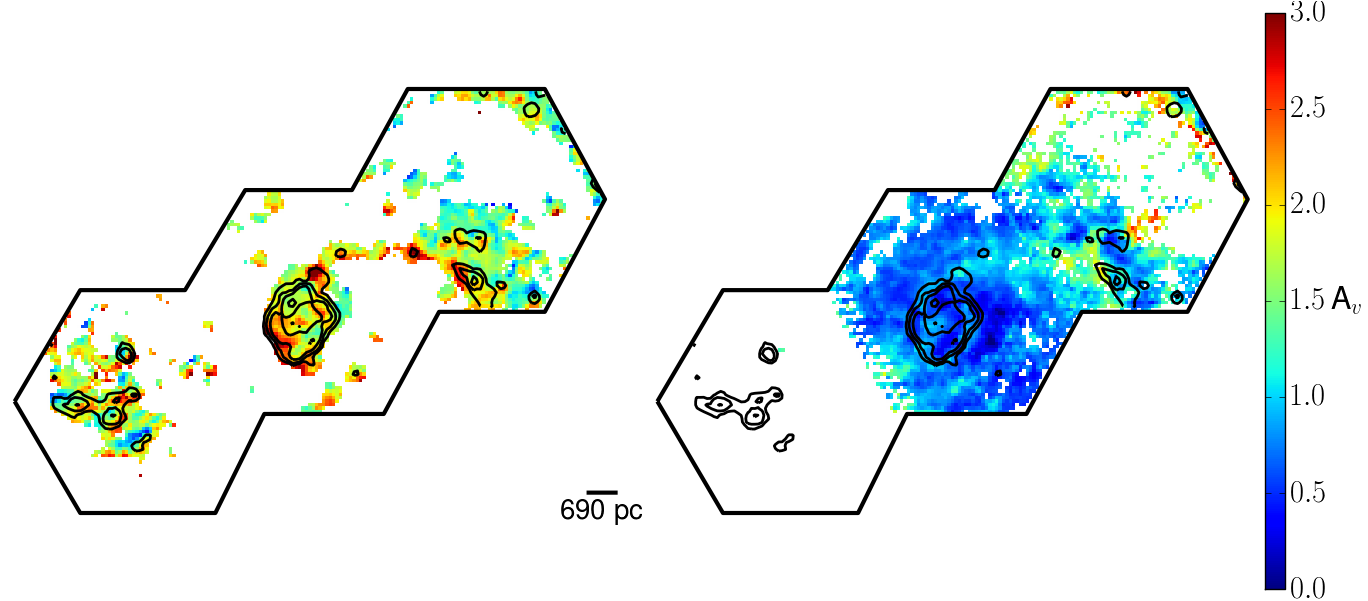}
\end{figure*}

\begin{figure*}[ht!]
\centering
\includegraphics[height=1.8in]{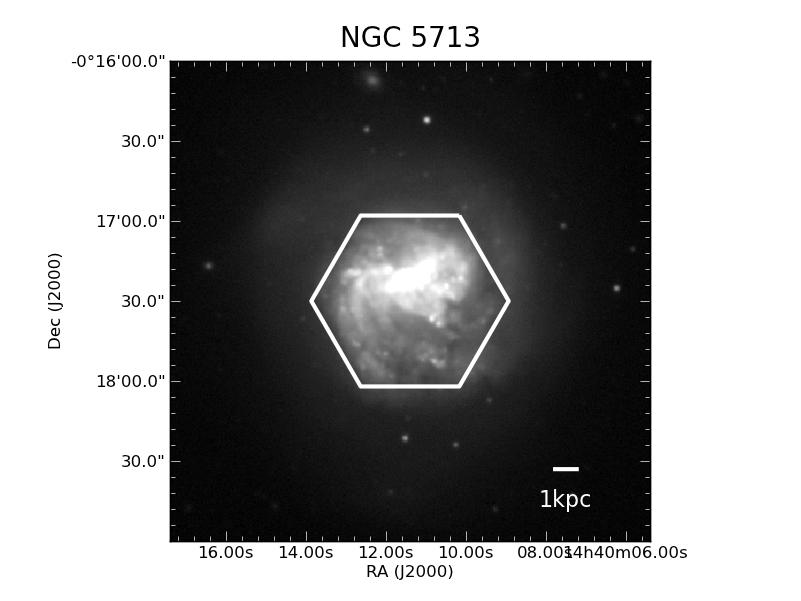}
\hspace{-.5cm}\includegraphics[height=1.6in]{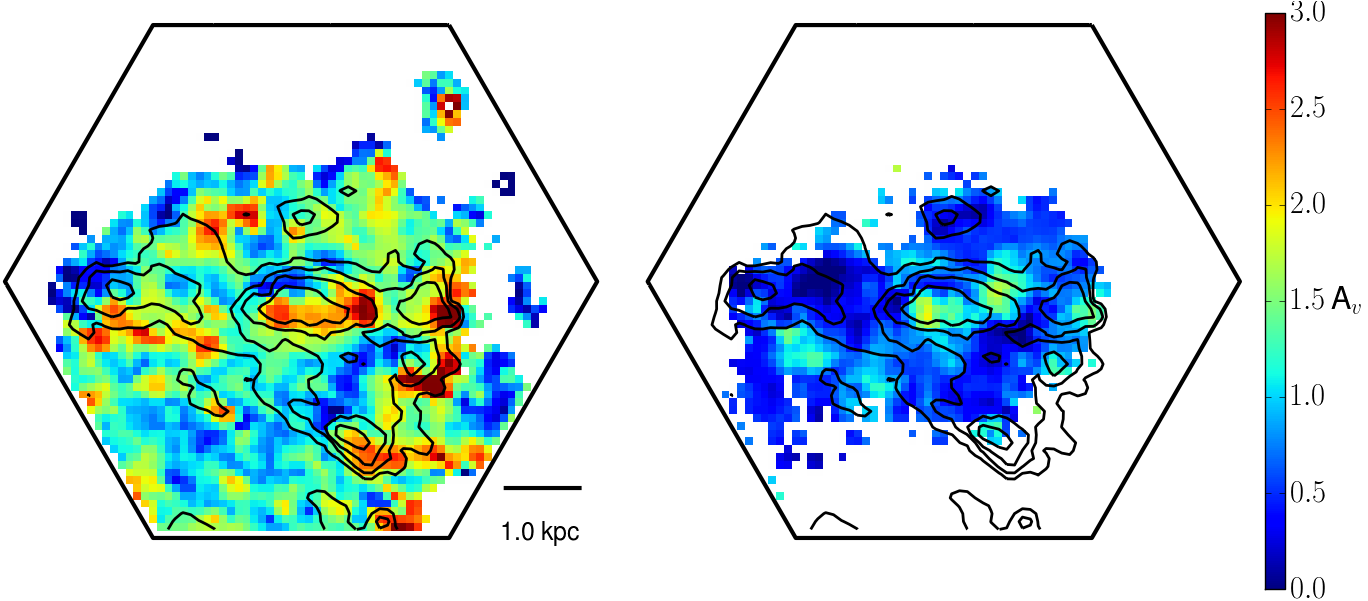}
\includegraphics[height=1.8in]{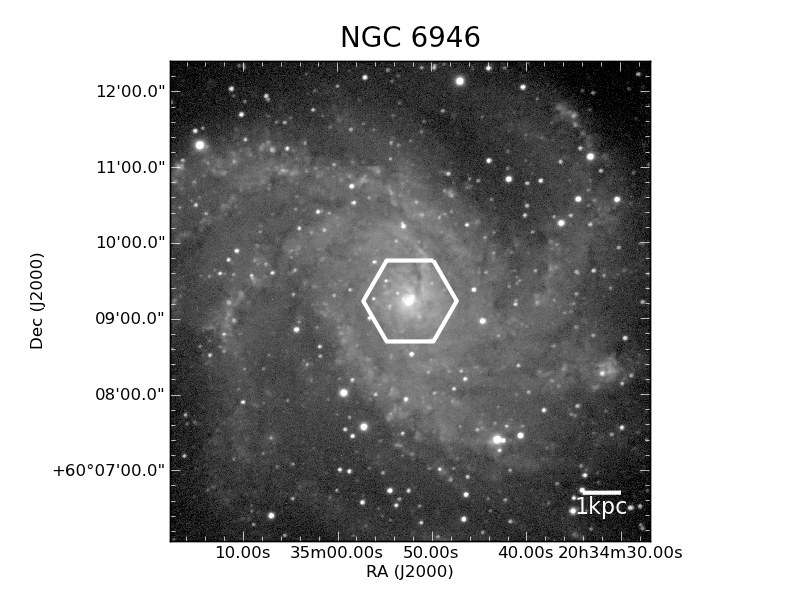}
\hspace{-.5cm}\includegraphics[height=1.6in]{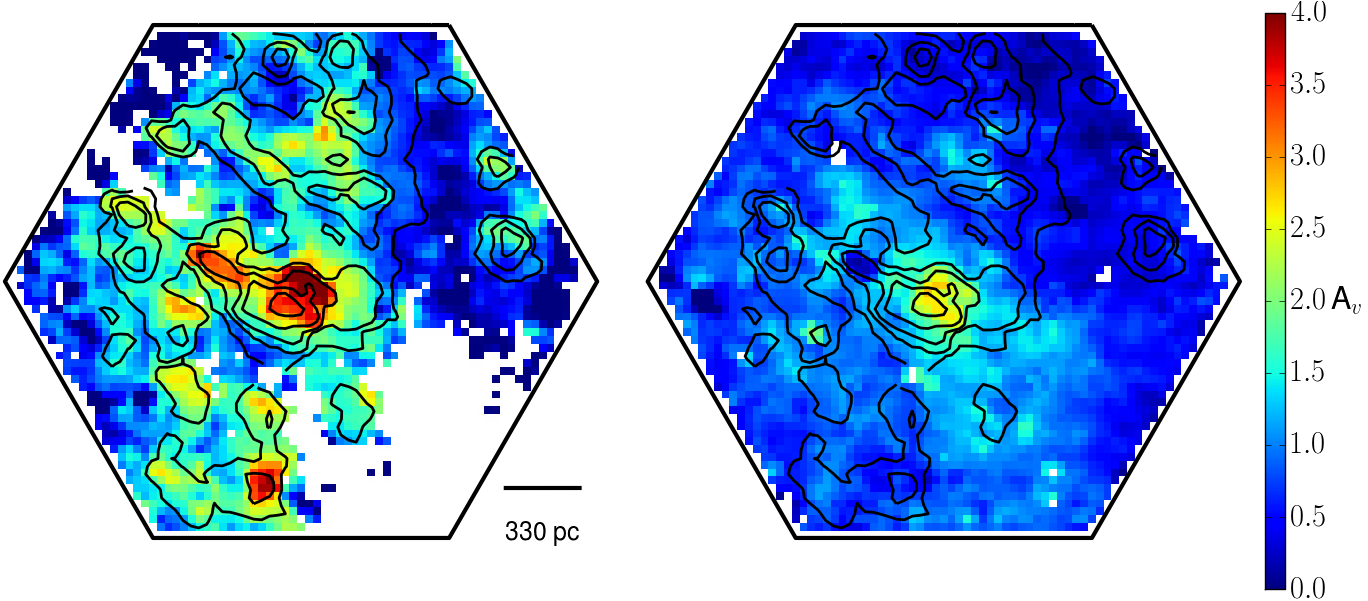} 
\includegraphics[height=1.8in]{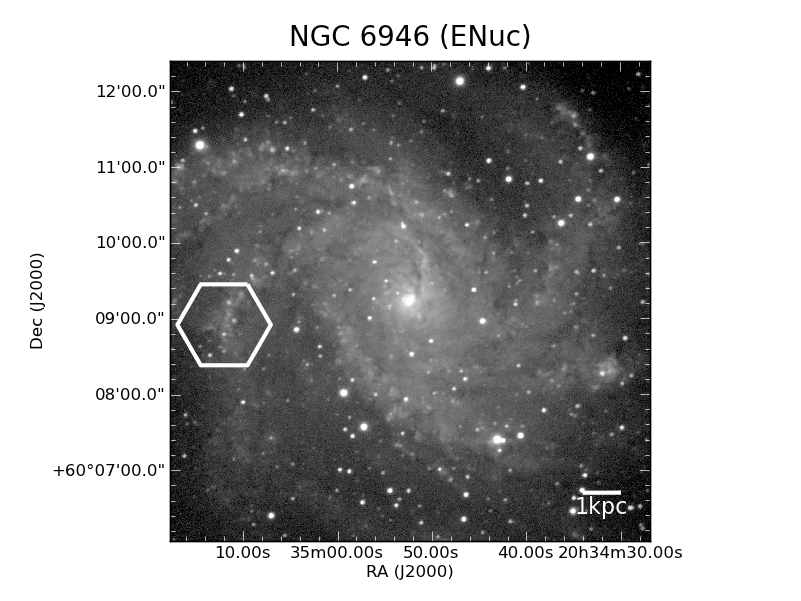}
\hspace{-.5cm}\includegraphics[height=1.6in]{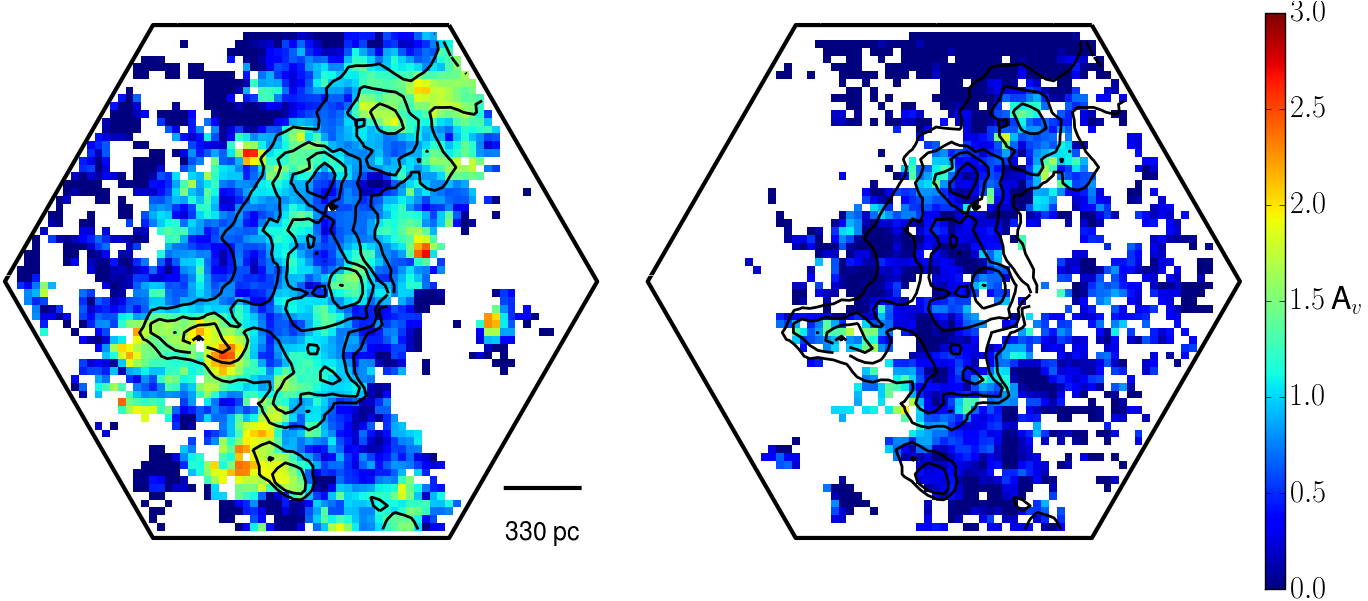}
\includegraphics[height=1.8in]{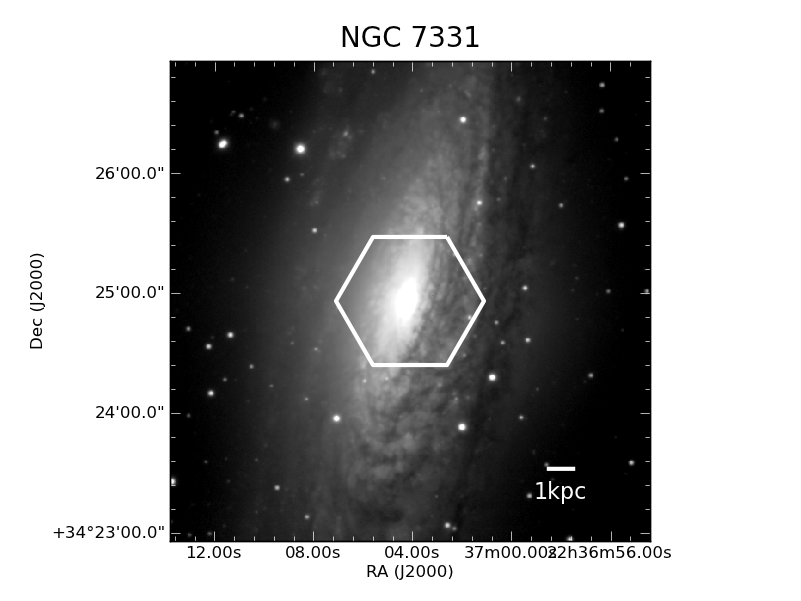}
\hspace{-.5cm}\includegraphics[height=1.6in]{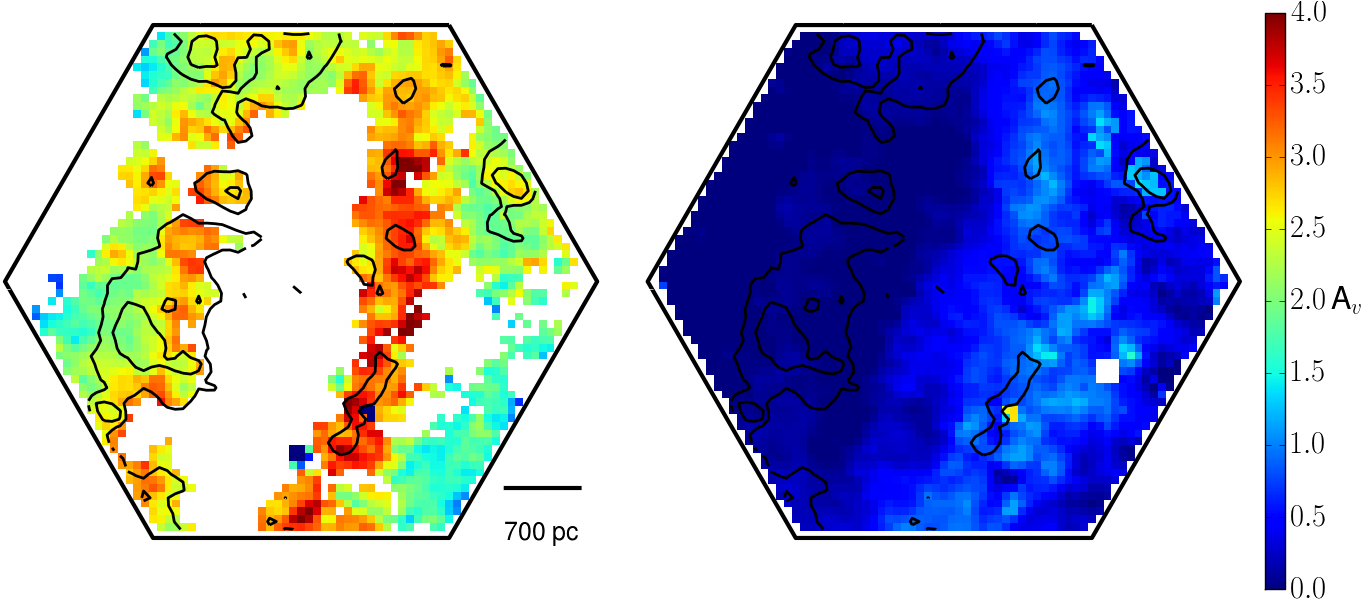}
\caption{Left: R-band images of each galaxy with the hexagonally shaped PPAK field of view indicated for each pointing. Center: A$_{V,g}$ maps from PPAK data based on the extinction in the Balmer decrement.  Right: A$_{V,s}$ maps from PPAK data matching the polynomial scaling component in the stellar template fits to a \cite{Calzetti2000} extinction law curve.  For PPAK maps a 10\arcsec\ scale bar is shown in each figure, with the corresponding physical scale listed underneath.
Optical images are taken from the SINGS DR4 \citep{SINGS} except for NGC 2146, which was observed on the KPNO 0.9m telescope \citep{Cheng1997}, and NGC 3077, which was observed on the KPNO 2.3m Bok Telescope \citep{Dale2009}.  Blanked regions in PPAK images show a signal to noise cut at positions where the amplitude in the line emission was less than three times the amplitude of the residual noise, or a  signal to noise cut of 10 at each position for the stellar continuum spectra.  Contours show the position of H$\alpha$ emission as detected in our data set. Corresponding A$_V$ maps are shown with the same color scale, as shown on the right.\label{fig:optical}\label{fig:avg}\label{fig:avs}}
\end{figure*}

We use the V300 grating, which covers 3700\AA~ to 7000\AA~ at $\sim$200 km s$^{-1}$ velocity resolution, and allows us to use a region of the CCD minimally affected by vignetting \citep{MarmolQueralto2011}. 
Each pointing was observed with three dither positions, shifted by (+1\farcs56,+0\farcs78) and (+1\farcs56,-0\farcs78) relative to the initial position, to completely recover the flux. Exposure times are listed in Table \ref{tab:ppak}. At each science target pointing, calibration lamp images were obtained to allow us to trace the positions of the spectra across the CCD, and He+HgCd arc lamp images were obtained to provide wavelength calibration. Twilight flats were obtained for accurate flat fielding of the CCD, and spectrophotometric standard stars were observed at the beginning and end of each night for flux calibration.

\subsubsection{Reduction} 
All PMAS/PPAK data are reduced using the p3d package \citep{Sandin2010}, version 2.1.2. All frames are bias subtracted, and observations from 2012 are median combined.  Observations from 2011 are cleaned of cosmic rays following the L.A. Cosmic technique \citep{vanDokkum2001}, as adapted for IFS usage within p3d, using a one-dimensional median filter with 8$\sigma$ clipping.  To ensure that strong emission line features are not identified as cosmic ray hits, we employ a 50$\sigma$ cut for arc lamp and science images. To determine the position of each spectrum on the CCD, peak emission in the calibration lamp images is traced along the length of the dispersion direction to determine a trace center.  Due to the close proximity of fibers on the CCD, a modified optimal extraction (MOX) method is employed that simultaneously fits all line profiles with a Gaussian function \citep{Horne1986}.  This method has been shown to reduce contamination by crosstalk from emission in neighboring fibers to a few percent \citep{Sandin2010}.  In the final science spectra extraction we allow for a slight shift in the trace centers, to correct for flexture in the instrument between observations, keeping the shape and other extraction parameters fixed. 

The He+HgCd calibration lamp images are used to provide wavelength calibration accurate to an rms of 0.3~\AA.  A small uniform offset in the final wavelength solution is allowed for each science target to account for flexture within the instrument.  This offset is determined by centering the brightest telluric lines (5577~\AA, 6300~\AA, 6863~\AA) present in each spectrum.  Sky flat and science target images are additionally corrected for a scattered light contribution by  first masking all regions of the data in close proximity of the trace centers, smoothing over unmasked regions using a median kernel, interpolating to masked regions using a seventh order polynomial along the cross-dispersion axis, and finally subtracting the resulting scattered light estimation.  Uniform illumination of the CCD within an IFS for flat fielding is difficult, and the PMAS PPAK instrument is further strongly affected by vignetting at the corners of the CCD \citep{MarmolQueralto2011, Sanchez2012}.  We use the min/max-filtered average of the twilight sky images to determine both the correction for the relative throughput between fibers as well as a smooth variation in the background level as a function of wavelength for each fiber, and multiply all science spectra by the resulting fiber-based flat field.  

Relative flux calibration is performed using light from a single fiber from observations of the standard stars BD +33d2642, BD +25d4655, HZ 4 and GRW +70d5824 only when they are well centered within that single fiber.  Absolute flux calibration is applied through comparison with existing photometrically calibrated broadband images.  As pointed out by \cite{Munoz-mateos2009}, there are non-negligable zero-point offsets in the SINGS DR4 optical images \citep{SINGS} that we find can affect the flux scaling by up to 30\%, so we choose instead to use Sloan Digital Sky Survey III (SDSS, \citealt{Ahn2012}) r-band images.  For these extended nearby galaxies, mosaics were created using the online service\footnote{http://data.sdss3.org/mosaics} that both combines and background subtracts neighboring imaging fields \citep{Blanton2011}.  The flux scaling is determined by measuring the r-band magnitude for all fiber positions in the optical images and scaling the image cube by the median offset.  Scalings determined from the g-band agree within 5\%, but we use here only the $r-$band calibrations to avoid the larger uncertanties at shorter wavelengths.  Comparison between the calibration of the data using SDSS as compared to SINGS that we apply here agrees to within 10\% with the recalibration given by \cite{Munoz-mateos2009}.  As conditions were photometric and scaling factors varry by less than 5\% throughout each night, we apply the average scaling factor from the corresponding observing night to NGC 2146 and NGC 6946, for which no SDSS photometry is available.  Comparing an integrated region from our final data cube with optical spectra from the central 20\arcsec\ $\times$ 20\arcsec\ region obtained by the SINGS project \citep{Moustakas2010} for a subsample of our targets suggest that our absolute flux calibration is accurate to $\sim$10\%, though it does show some evidence of a residual gradient in the continuum flux at about the 5\% level, where our spectra are redder.  These errors are within the estimated errors as propagated through the p3d pipeline.  We find $\sim$15\% errors throughout most of our observed spectrum and significantly higher errors (up to 50\%) in the low sensitivity blue end of the spectrum.  

The result of the reduction process is a row-stacked spectrum (RSS) file, in which each of the 331 spectra for each dither are stored individually. A separate position table maps each spectrum onto the image plane.  The fiber positions are almost entirely regular, however slight deviations are visible in the outer part of the field of view.  We use the fiber positions provided by Martinsson (2012, PhD Thesis), which were measured from a high resolution image of the instrument.  To ease visualization and comparison with data at other wavelengths, we grid these RSS files from all dither positions together onto a regular 1$^{\prime\prime} \times 1^{\prime\prime}$ grid using a Delaunay linear triangulation individually for each wavelength to construct an image cube.  Mosaiced pointings in NGC 3627 and NGC 4321 are gridded together into a single image cube.  Comparison of spectra from overlapping regions in these two galaxies show agreement to within 10\% throughout most of the wavelength range, and to within $\sim$20\% below 4000~\AA, in good agreement with our estimated errors.  As all observations were taken at airmasses below 1.5 we omit any corrections for differential atmospheric refraction, which is expected to cause only subfiber ($<$ 1\arcsec) shifts across the wavelength range in all galaxies.  Tests with dithered observations of standard stars and with simulated PPAK data reveal variations in the shape of the effective point spread function (PSF) depending on where the point source falls within and between fibers, with effective FWHM value ranging from 1\farcs5 to 3\farcs5 and a median of 2\farcs5.  

To remove the sky contamination, integrated spectra from each dither position  with all galactic emission lines masked are first fit with a combination of stellar templates and the median of each dedicated sky field observed throughout the night.  The resulting linear combination of sky templates is then subtracted off of each spectrum in the image cube.  This is found to correctly remove the bulk of the sky emission features, but is ineffective at removing the strongest OH emission features at 5700, 6000 and 6300~\AA~ which are particularly variable.  These lines are masked throughout the rest of our analysis.  We also divide each spectrum by a multiplicative contribution of foreground Milky Way extinction based on the \cite{Cardelli1989} extinction law using the Milky Way foreground reddening determined by \cite{Schlafly2011} revisions to the \cite{Schlegel1998} dust maps at each galaxy position.

\subsubsection{Line Maps} 
\label{sec:linemaps}
The resulting image cubes from the optical IFS data allow us to construct maps of the line emission from H$\alpha$ and H$\beta$, for which we respectively achieve typical 3$\sigma$ flux sensitivities of $5 \times 10^{-16}$ erg s$^{-1}$ cm$^{-2}$ arcsec$^{-2}$ and $1 \times 10^{-16}$ erg s$^{-1}$ cm$^{-2}$ arcsec$^{-2}$.  Each gridded spectrum is fitted using the GANDALF software package \citep{Sarzi2006}, which employs penalized pixel-fitting (pPXF, \citealt{Cappellari2004}) to allow simultaneous fitting of both the stellar continuum and a full suite of optical emission lines from [O\textsc{ii}] 3727\AA~ to [S\textsc{ii}] 6731\AA.   For each spectrum, we fit a linear combination of simple stellar population (SSP) template spectra taken from the \cite{Tremonti2004} library of \cite{Bruzual2003} templates for a range of ages (5 Myr to 11 Gyr), resampled from an initial 3\AA~ resolution to match our observed $\sim$7\AA~ FWHM resolution.  We also considered including templates with a constant or an exponentially declining star formation history, however as these were generally not selected by the fitting software we have omitted them from this analysis. 
In all cases we restrict the metallicity to be solar due to the limited range in our sample (see Table \ref{tab:properties}).  

\begin{figure*}
\centering
\includegraphics[width=6.2in]{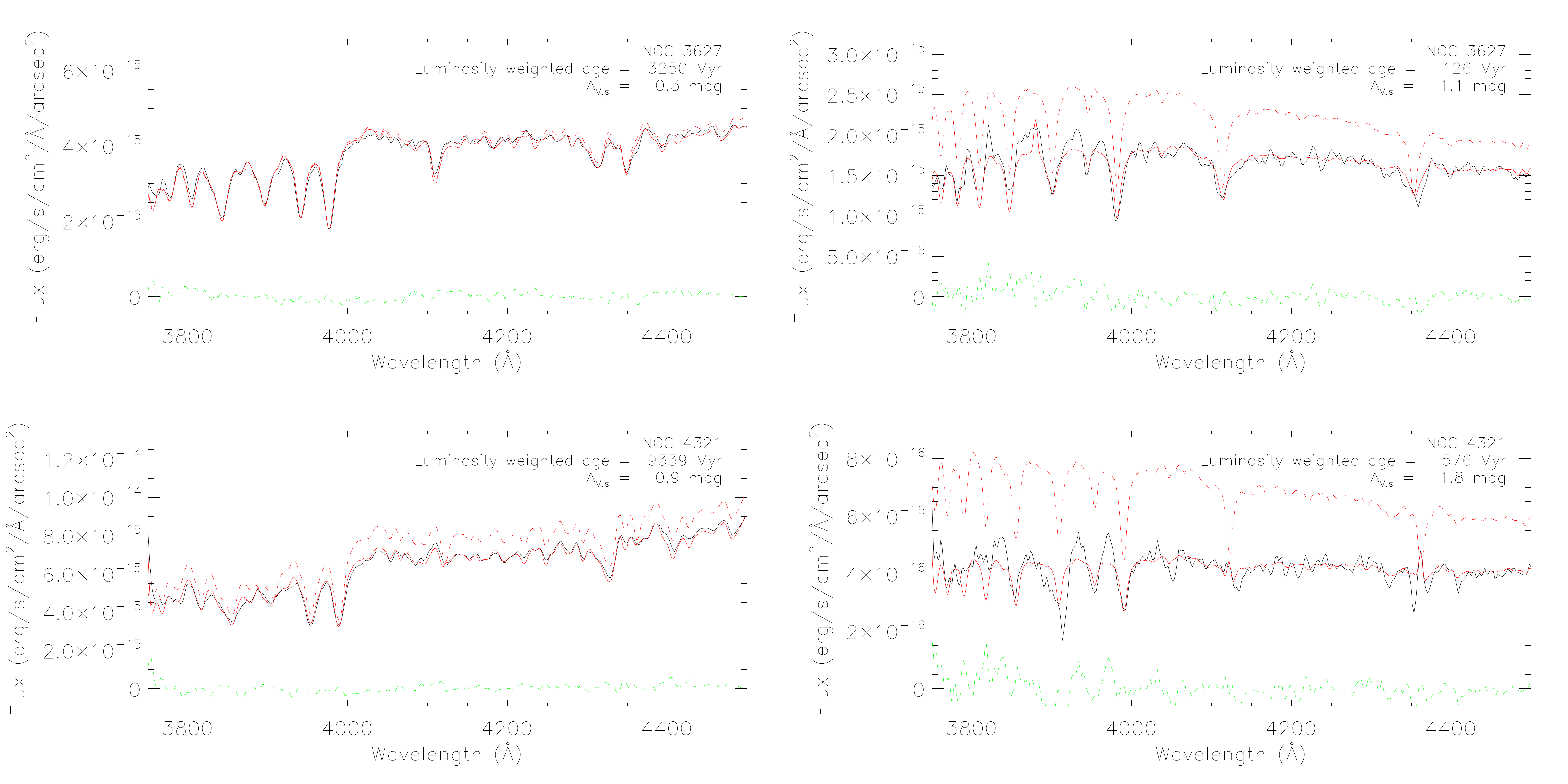}
\caption{Sample observed spectra demonstrating GANDALF fits breaking the age-extinction degeneracy.  The observed spectra (black) and resulting GANDALF fit (solid red) are well matched, with residuals shown below (green).  Also shown is the combination of unreddened stellar templates used in the fit (dashed red). Selected spectra show young and old stellar populations in NGC 3627 and NGC 4321 that have varying levels of extinction.   Age sensitive features, like the D4000 break and high-order Balmer absortion lines H$\delta$, are well fit in all cases.
\label{fig:samplefits}}
\end{figure*}

We additionally allow for a multiplicative third order Legendre polynomial correction to the stellar templates.  This functionallity is included in the pPXF software package to correct for low frequency continuum variations, and we find that in practice it is well matched to the \cite{Calzetti2000} attenuation law and can be used to measure reddening of the stellar continuum by dust attenuation (see Section \ref{sec:constructing}). Templates have an equivalent width of the stellar absorption at H$\alpha$ that is 1-2\AA.  The selection of templates appears well-behaved, as neighboring regions generally show smooth transitions in the total luminosity-weighted age (see also Section \ref{sec:absorption}).  The age-extinction degeneracy is broken through the simultaneous SSP template fitting and polynomial reddening correction (Figure \ref{fig:samplefits}), with age sensitive features like the 4000\AA~ break and high-order Balmer absortion lines well fit in all cases.

We perform this GANDALF analysis on each spectrum of our full 2\farcs5 resolution image cubes.  In addition, our data allow us the flexibility to match resolution with the lower resolution dust mass surface density maps (see Section \ref{sec:pixbypix}).  Convolutions are done using the kernels and techniques described in \cite{Aniano2011} to convert our effective 2\farcs5 Gaussian PSF to the 18\arcsec\ SPIRE 250 $\mu$m PSF.  We neglect potential variations in the PSF shape within the IFS field of view as our final convolved resolution is sufficiently degraded that it is relatively insensitive to those differences. 
We perform this convolution for each spectral channel of the optical image cube and interpolate to match the WCS pixel grid of the dust mass surface density maps.  Then we re-run our spectrum fitting software on each pixel of the resulting cube to construct line maps at matched 18\arcsec\ resolution.  This improves the signal to noise and also more accurately reflects the luminosity-weighting effects when calculating further derived quantities.     

\section{Results} 
\label{sec:results}

\subsection{Constructing A$_V$ maps} 
\label{sec:constructing}

With the resulting line maps we measure the attenuation of the nebular lines, A$_{V,g}$, by the Balmer decrement, under the assumption of case B recombination and an electron temperature of $10,000$ K (giving an intrinsic H$\alpha$/H$\beta$ = 2.86),   as 
\begin{equation}
\label{eqn:1}
E(B-V)_g = \frac{2.5}{k(H\beta) - k(H\alpha)} \log_{10} \left(\frac{F_{H\alpha}/F_{H\beta}}{2.86}\right) 
\end{equation}
\begin{equation}
\label{eqn:2}
A_{V,g} = R_V~ E(B-V)_g
\end{equation}
Here, F$_{H\alpha}$ and F$_{H\beta}$ are the measured line fluxes, and we assume $k(\lambda)$ from the \cite{Calzetti2000} attenuation law and a typical Milky Way value of  R$_V$ = 3.1, resulting in values for $k$(H$\alpha$) and  $k$(H$\beta$) of 2.38 and 3.65, respectively. Variation in the electron temperature by a factor of two would result in $\sim$0.1 magnitude difference in the A$_{V}$ calculated.  A gradient in the relative flux calibration of 5\% would affect the A$_{V}$ estimates by less than 0.1 magnitudes.

\begin{figure}[b!]
\centering
\includegraphics[width=3.2in]{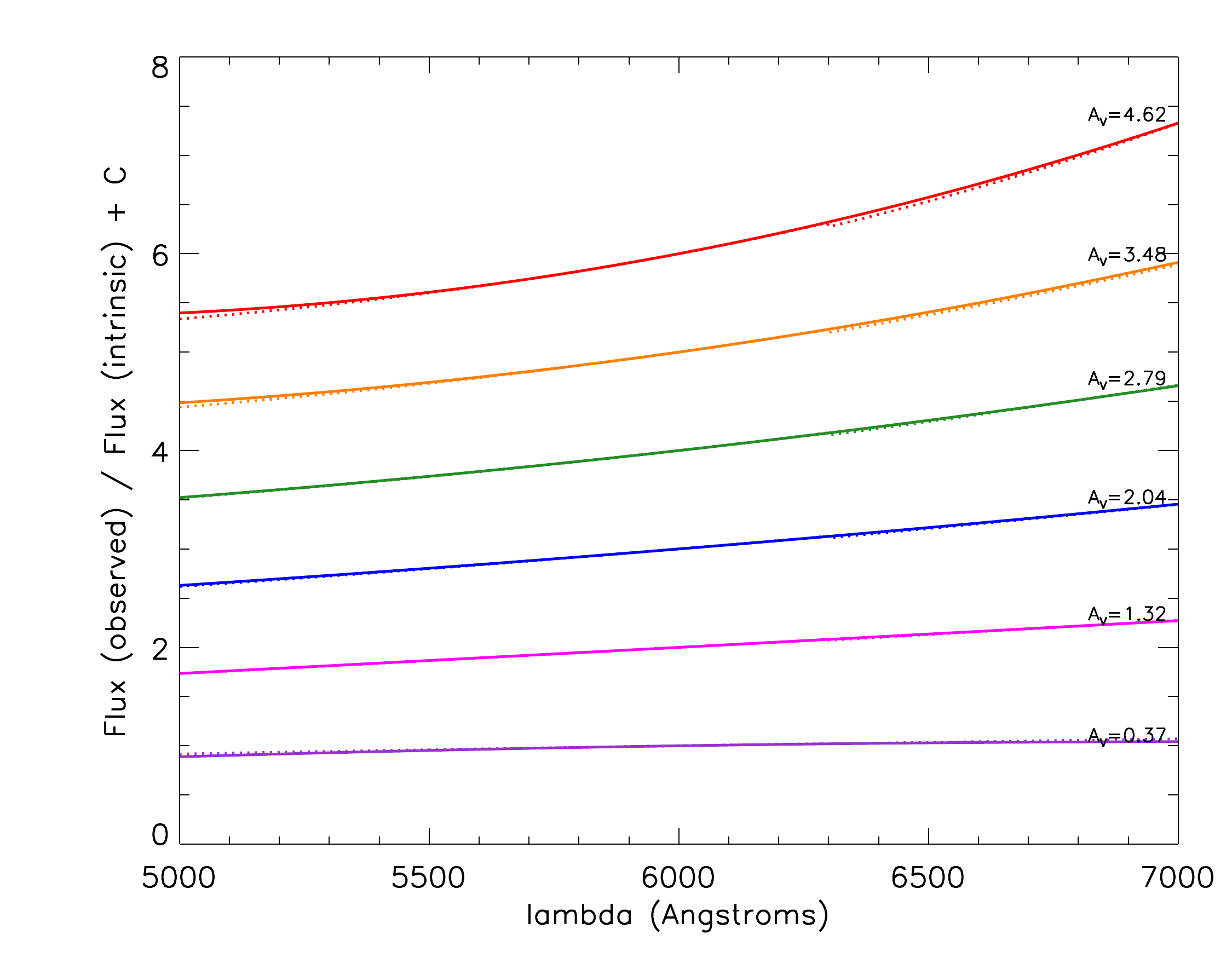}
\caption{Sample multiplicative polynomials (solid lines) and corresponding fit reddening curve (dotted lines).  Only the curve between 5000-7000 \AA~ is considered, and only a linear polynomial fit giving the slope is matched.  All curves are taken from NGC 2146, which shows the widest range of stellar reddenings due to a high inclination and strong dust lane feature.    \label{fig:starav_fit}}
\end{figure}

The convolution we performed on the optical image cubes to match the lower resolution dust mass surface density maps (see Section \ref{sec:linemaps}) treats blanked low signal to noise regions and edges in the line maps as pixels to interpolate over, then replaces them with NANs after the convolution.  This is equivalent to the assumption that the edge emission continues uniformly beyond the map region, which given the clumpy nature of H\textsc{ii} regions may not be true.  This can lead to larger uncertainties in regions surrounding missing pixels and map edges, as data outside the region are not convolved into the optical line maps though it is included in the larger field of view dust mass maps. We test these edge effects on our convolution by combining existing continuum subtracted H$\alpha$ maps \citep{SINGS}  with our high resolution H$\alpha$ line maps.  Assuming a fixed A$_V$ of 1 magnitude outside of our PPAK region, approximately the median A$_V$ we observe but more likely an overestimate given our detection limits, we use the H$\alpha$ images to fill in blanks in the H$\beta$ maps.  Convolving these images and re-measuring the Balmer decrement, we find variations of 10-20\% in the pixels immediately at the image edge or near regions of low H$\alpha$ or H$\beta$ signal to noise (i.e. NGC 7331 and NGC 3627 have no H$\alpha$ emission in the center), but significantly smaller variations (less than 5\%) in the rest of the field.  These uncertainties lead to absolute errors of around 0.1 magnitudes at the field edge, though we expect this to be an overestimate for the less extended galaxies.  We have factored these into the error estimates for each field.

Additionally, assuming the multiplicative third order Legendre polynomial used to scale the stellar SSP templates is due to the internal reddening curve of the target galaxy, we match the linear slope across the 5000-7000 \AA~ wavelength range to the slope of the \cite{Calzetti2000} law attenuation curve to determine A$_{V,s}$, the V-band extinction measured from reddening of the stellar continuum light.  Normalizing all curves at 6000 \AA, the match is reliable for a range of extinctions (Figure \ref{fig:starav_fit}), with typical errors of 0.1 magnitude. Deviations for the highest extinction regions are due to the low signal to noise of the stellar continuum light given the extreme dust extinction, and particularly on the blue end the residual calibration uncertainties become apparent.  

\subsection{Comparison of emission line and stellar continuum A$_V$ maps}  
The A$_V$ maps determined by the Balmer decrement and from the stellar continuum reddening show qualitatively very similar features on 100-300 pc scales (Figure \ref{fig:avs}). Blanked regions indicate locations with low signal to noise, as in NGC 3627 and NGC 7331 where there is little H$\alpha$ emission detected in the center.  Extinction maps generally follow the strong dust lane features apparent in the optical images, particularly in NGC 2146 and NGC 7331.  In most galaxies the regions of higher extinction coincide with regions of higher H$\alpha$ emission.   NGC 7331 shows different morphologies in the two A$_V$ maps, particularly as the A$_{V,s}$ maps shows relatively little extinction on the east side of the field.  Comparison with the optical images reveals a bright bulge, the light of which dominates the continuum  in the regions overlapping the far side of the disk.  We also note that the A$_{V,s}$ map for NGC 4321 on the east is omitted as those data were observed close to sunrise and the continuum light gradient is contaminated.  This does not affect our measurements of the line emission features.

\begin{figure*}[p]
\includegraphics[width=2.3in]{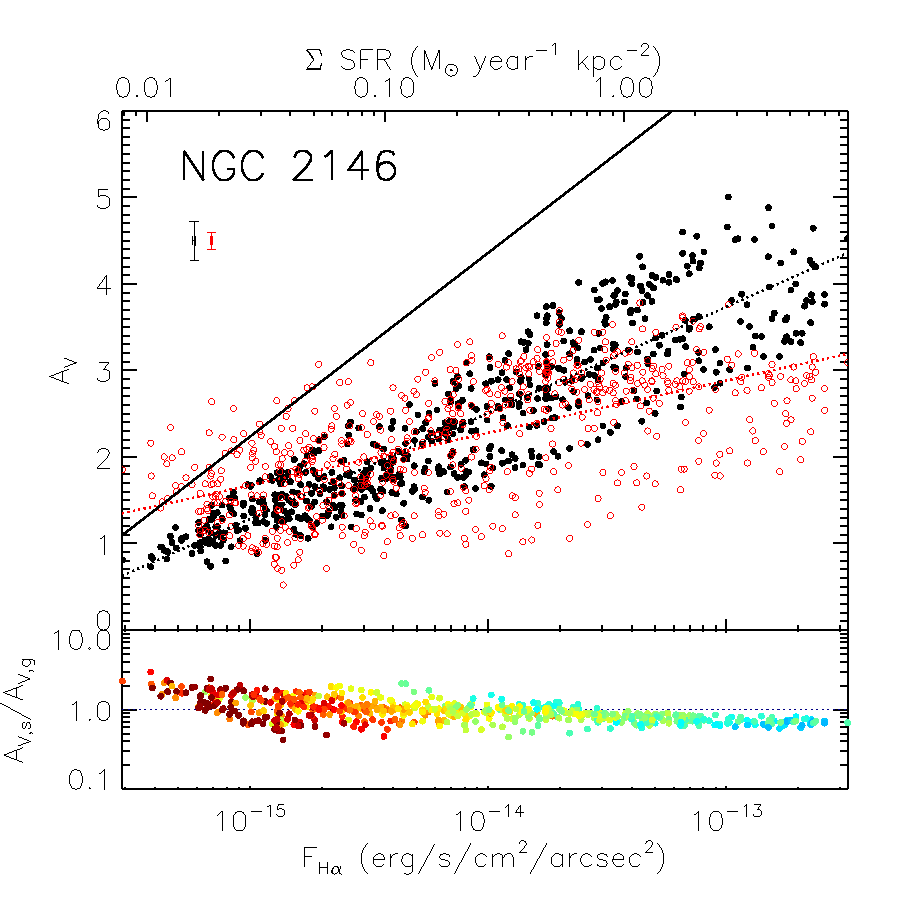}
\includegraphics[width=2.3in]{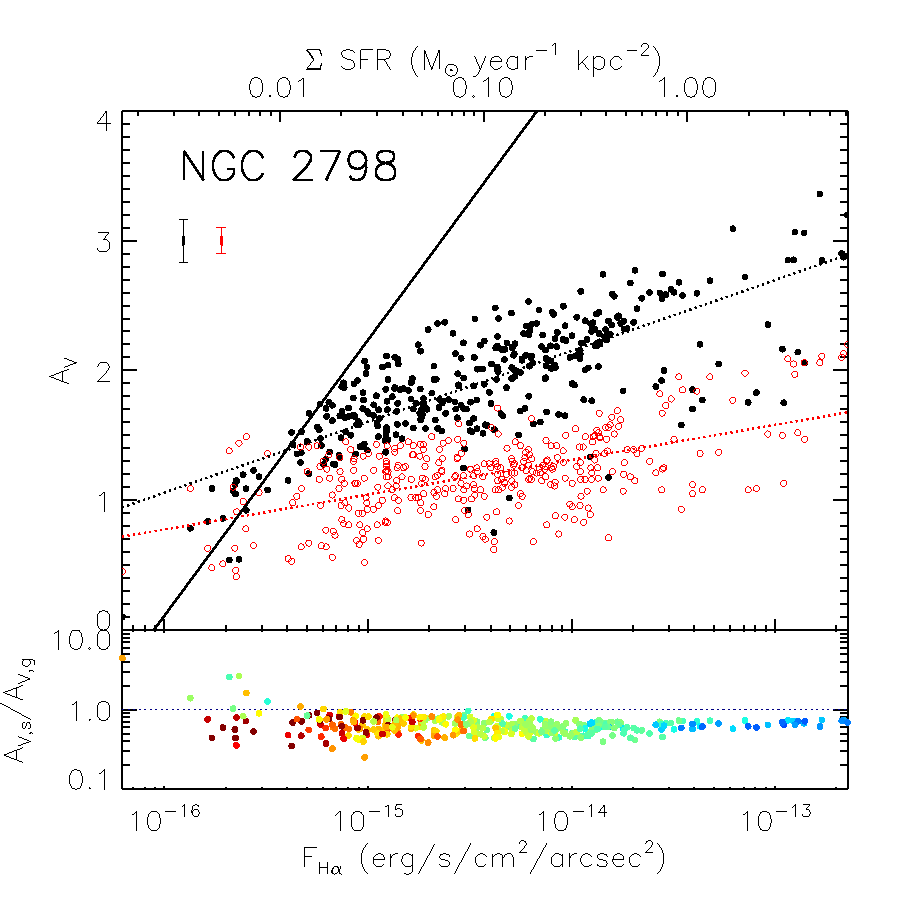}
\includegraphics[width=2.3in]{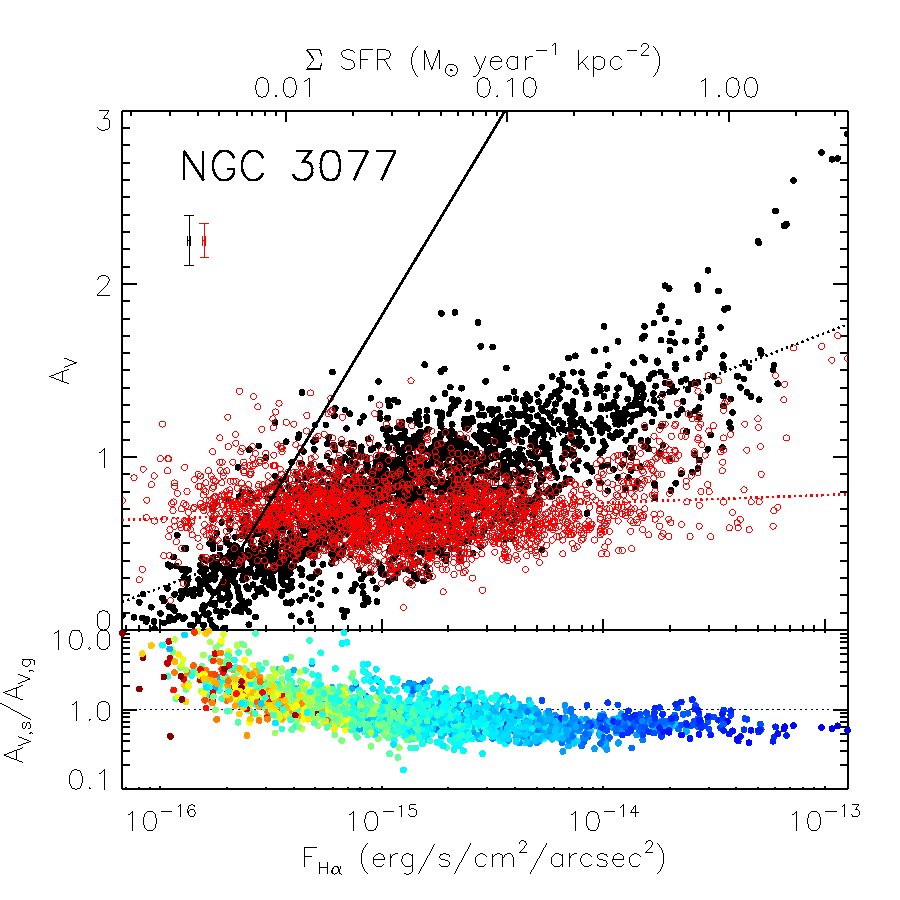}\\
\includegraphics[width=2.3in]{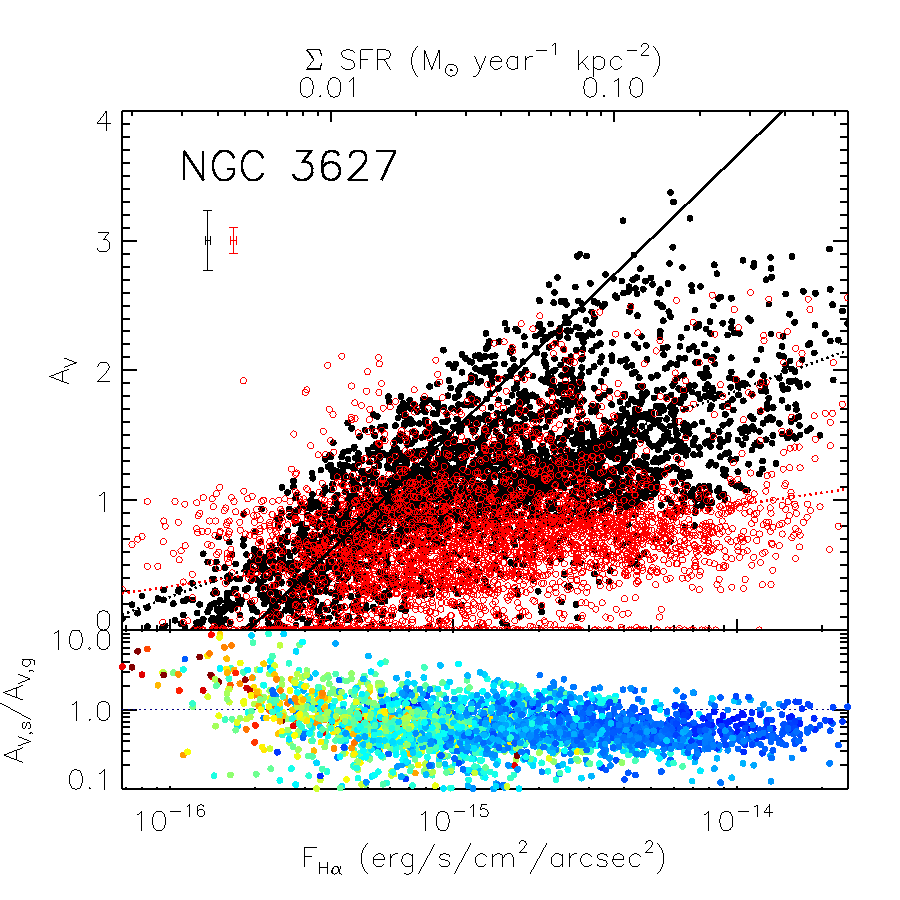} 
\includegraphics[width=2.3in]{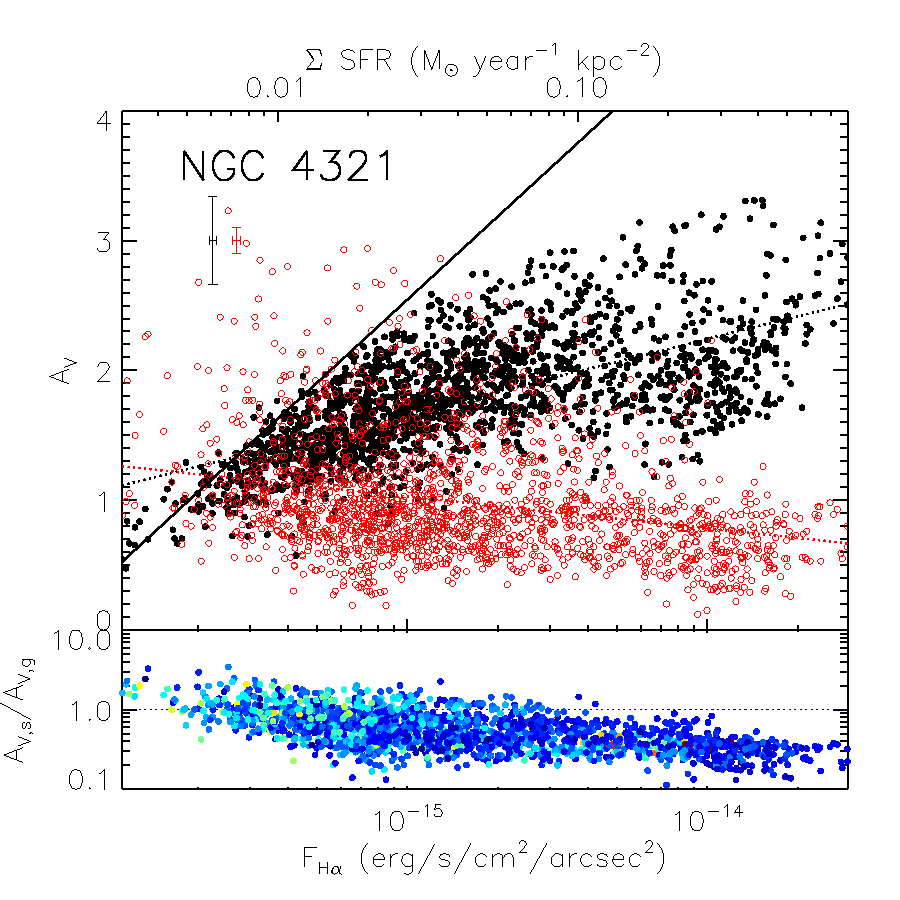}
\includegraphics[width=2.3in]{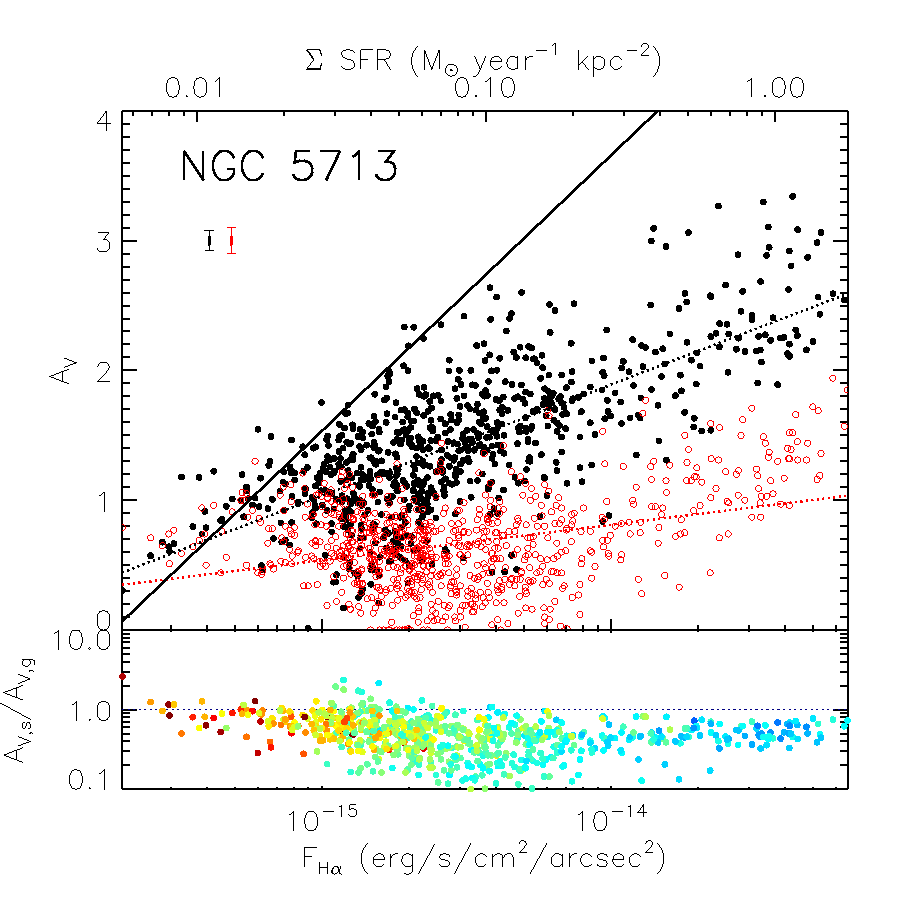}\\
\includegraphics[width=2.3in]{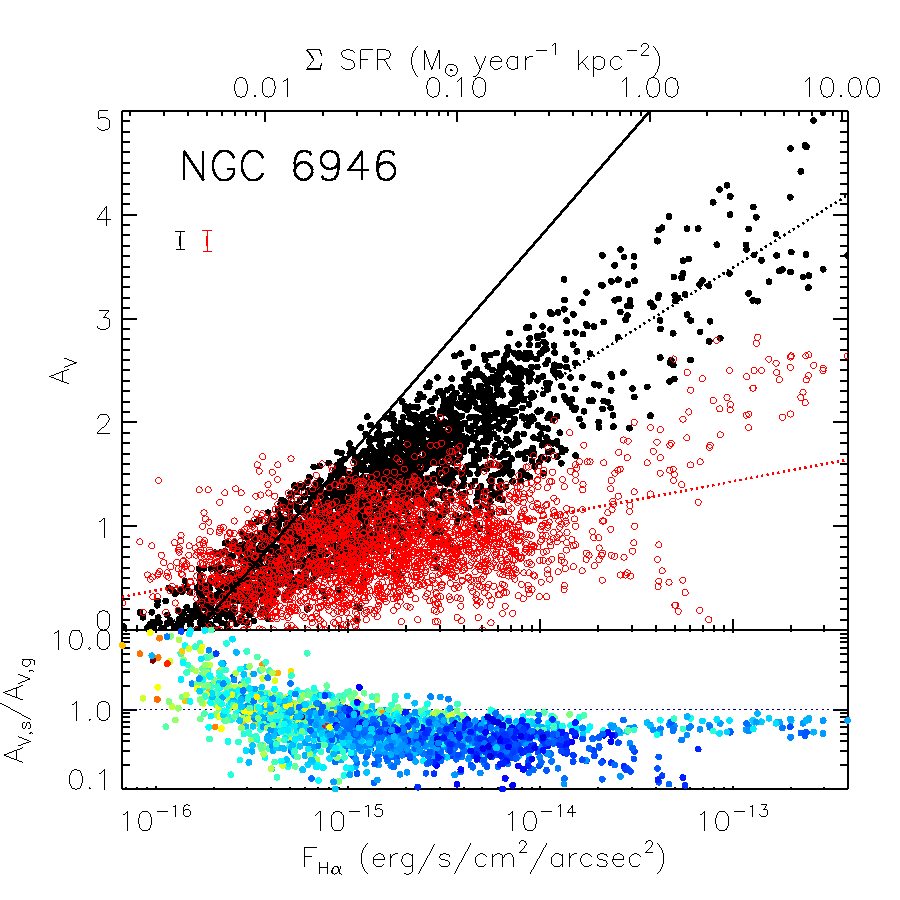}
\includegraphics[width=2.3in]{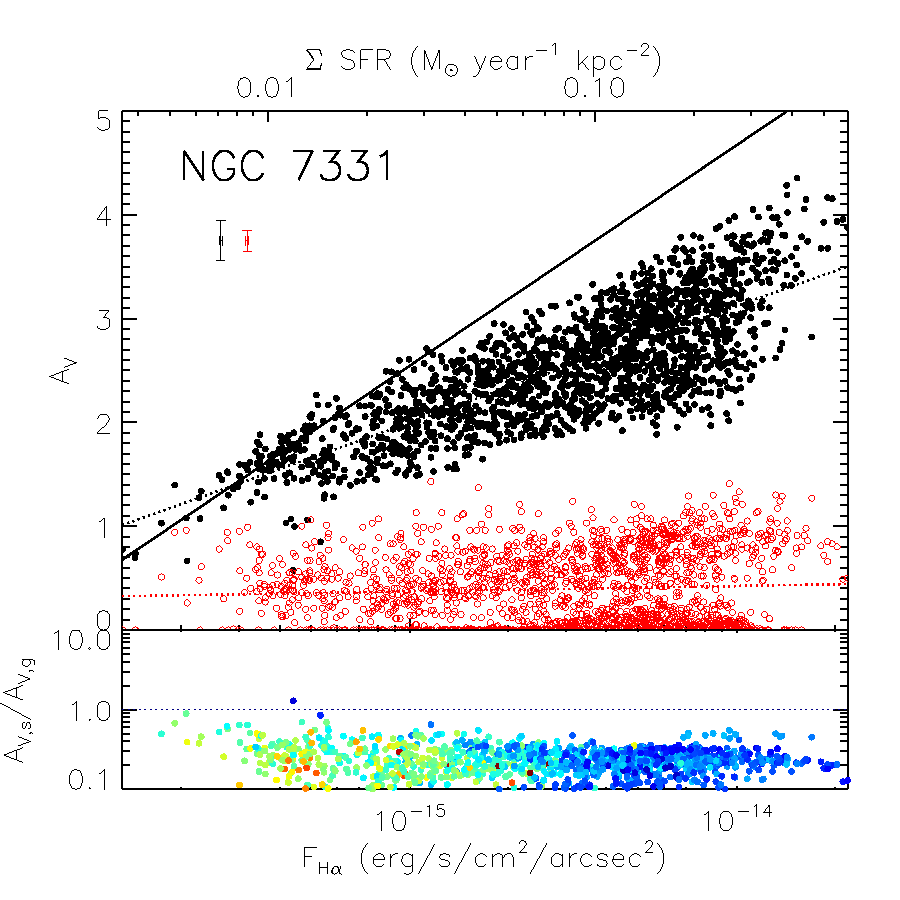}
\includegraphics[height=2.3in]{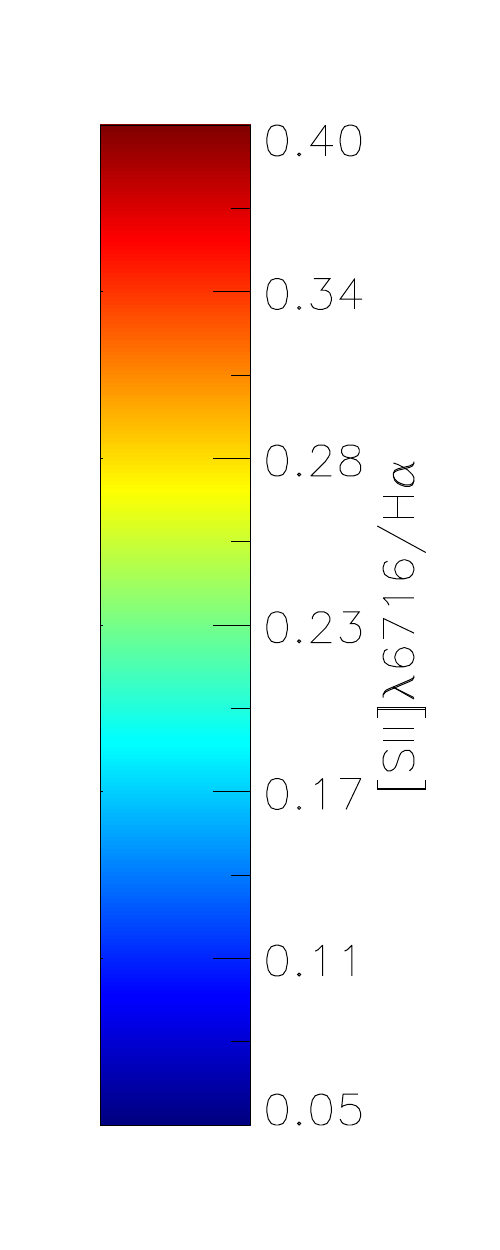}
\begin{center}
\caption{A$_V$ measured from the Balmer decrement (black) and stellar continuum extinction (red) as a function of extinction corrected H$\alpha$ flux for each of the eight galaxies.  Conversion to a $\Sigma_{SFR}$ is shown on the top axis.  The ratio of A$_{V,s}$ to A$_{V,g}$ is shown on the bottom panel of each plot, color coded by the [S\textsc{ii}]$\lambda$6716/H$\alpha$ ratio as shown in the colorbar on the lower right.  The solid line indicates the detection limit for A$_{V,g}$ based on the H$\beta$ sensitivity.  This detection limit does not apply to the measurement of A$_{V,s}$ from the stellar continuum reddening.  Dotted lines show a bisector fit to each of the A$_V$ measures. Only points with high signal to noise in both A$_{V,s}$ and A$_{V,g}$ are shown.  Sample error bars are shown in the upper left corner for each. 
\label{fig:sfrsd}}
\end{center}
\end{figure*}

More detailed comparison of the two A$_V$ measures show systematic differences.  Figure \ref{fig:sfrsd} shows a pixel by pixel comparison of A$_{V,g}$ and A$_{V,s}$ with the H$\alpha$ flux,  extinction corrected using A$_{V,g}$, at 2\farcs5 resolution for each of the eight galaxies.  The ratio of A$_{V,s}$ to A$_{V,g}$ is also shown, and is color coded by the [S\textsc{ii}]$\lambda$6716/H$\alpha$ ratio tracing the transition between H\textsc{ii} and diffuse ionized gas regions (see Section \ref{sec:absorption}).  Here we have excluded all low signal to noise points, taking only those where both the stellar continuum has S/N above 10 and the line fluxes have amplitudes more than 10 times the residual noise. We observe that A$_{V,g}$ is systematically higher than A$_{V,s}$, as previously reported by \cite{Calzetti1994,Calzetti1996,Calzetti2000}.  This has been attributed to the fact that line emission sources are preferentially located within dusty birth clouds \citep{Calzetti1994,Charlot2000}, whereas the stellar light mainly samples the diffuse dust in the ISM.  We see convincing evidence for this as none of our high SFR surface density ($\Sigma_{SFR}$) regions (above 1 M$_\sun$ yr$^{-1}$ kpc$^{-2}$ in our H$\alpha$ images) have A$_{V,g}$ below 1.  We have calculated the SFR surface density from our H$\alpha$ line maps, correcting for attenuation using the measured A$_{V,g}$ and assuming the \cite{Calzetti2007} conversion from H$\alpha$ luminosity to SFR,  
\begin{equation}
{\rm SFR} ({\rm M_\sun ~yr^{-1}}) = 5.3 \times 10^{-42} {\rm L}_{H\alpha} ({\rm erg~ s^{-1}}).
\end{equation}
There is also a general trend for low A$_{V,g}$ only in low $\Sigma_{SFR}$ regions (black dotted lines), whereas with A$_{V,s}$ values appear to trace less strongly the $\Sigma_{SFR}$ (red dotted lines).

We see in most galaxies a transition from  a consistently tight agreement in the ratio of A$_{V,s}$/A$_{V,g}$ at a value less than one in regions with high H$\alpha$ fluxes, to a ratio of approximately one with increased scatter at the lowest H$\alpha$ fluxes.  Trends towards a ratio greater than one are likely due to the detection limit of A$_{V,g}$ due to the sensitivity limits for the H$\beta$ flux, indicated as a solid line in Figure \ref{fig:sfrsd}, which does not affect $A_{V,s}$ as it is measured from the stellar continuum reddening but does limits detections to low values of A$_{V,g}$.  However given that we see no high $\Sigma_{SFR}$ regions with A$_{V,s}$/A$_{V,g} > 1$, and as the deviation is not within the errors, we believe this is convincing evidence of a transition from H$\alpha$ flux originating in buried H\textsc{ii} regions to H$\alpha$ flux instead originating from a more distributed morphology.  The diffuse ionized medium has been studied extensively in the Milky Way and in nearby galaxies \citep{Reynolds1984, Ferguson1996, Hoopes1996, Greenawalt1998, Thilker2002, Oey2007}, and is typically thought to result from a combination of ionizing radiation leaking from H\textsc{ii} regions and from shock heating.  Its distribution covers most of the star-forming disk though it is morphologically related to the H\textsc{ii} regions \citep{Wang1997}, and from our observations we conclude that it correlates with the dust distribution in a way similar to the stars  (see also the discussion in Section \ref{sec:absorption}).

\subsection{Pixel by pixel comparison of surface density of dust to A$_V$} 
\label{sec:pixbypix}

\begin{figure}[t!]
\centering
\includegraphics[height=1.8in]{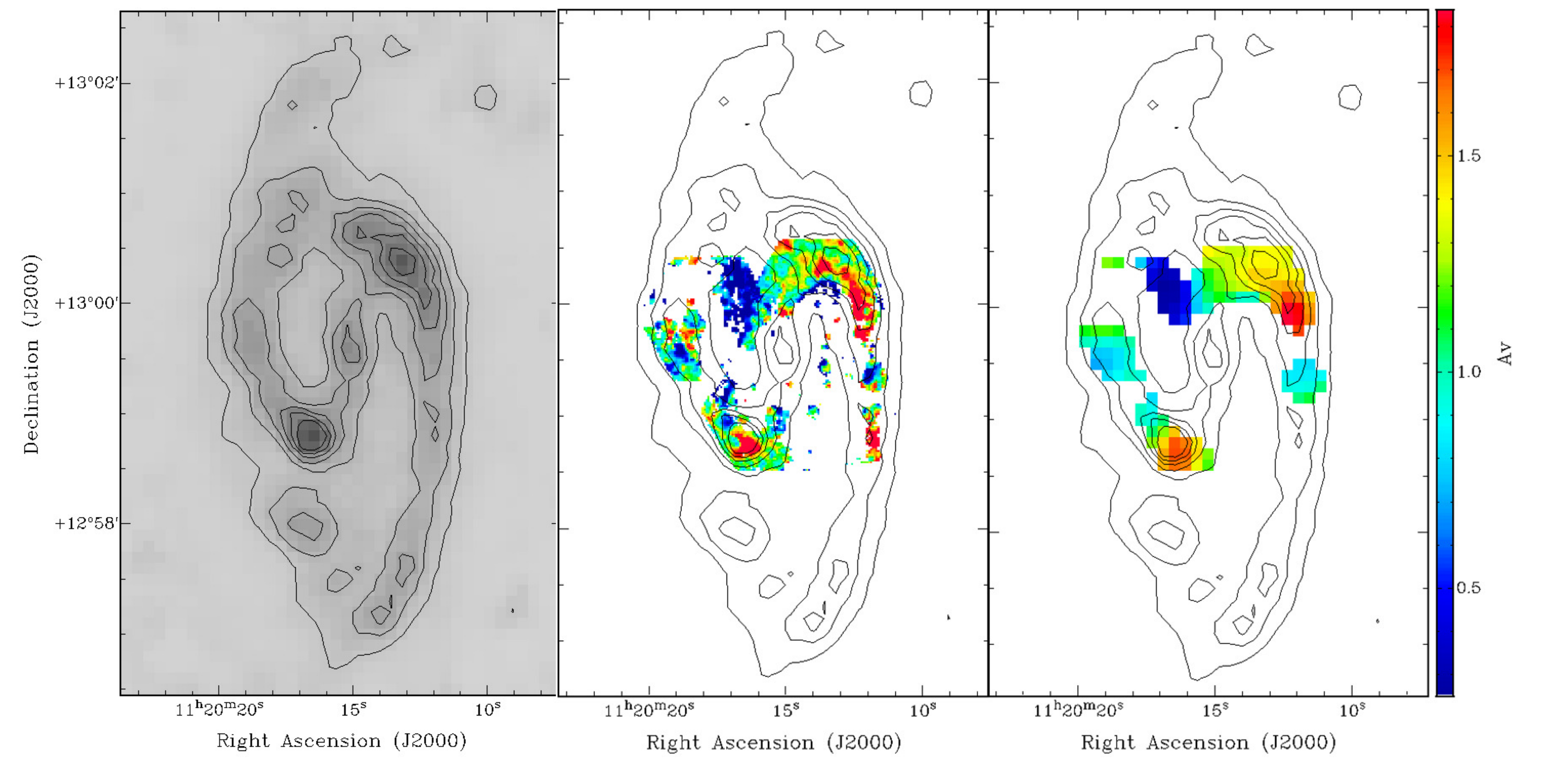}
\caption{Map of the $\Sigma$M$_d$ from the far-IR dust modeling (left) is compared to the A$_{V,g}$ maps at full 2\farcs5 resolution (center) and the convolved 18\arcsec\ resolution (right) for NGC 3627.  All contours are taken from the $\Sigma$M$_d$ maps with a lowest contour at $3 \times 10^5$ M$_\sun$/kpc$^2$ and linear spacing with the highest at $2 \times 10^6$ M$_\sun$/kpc$^2$.  Both A$_V$ maps are shown with the same color scale.
\label{fig:ngc3627}}
\end{figure}

\begin{figure*}[Hb!]
\centering
\includegraphics[width=6.5in]{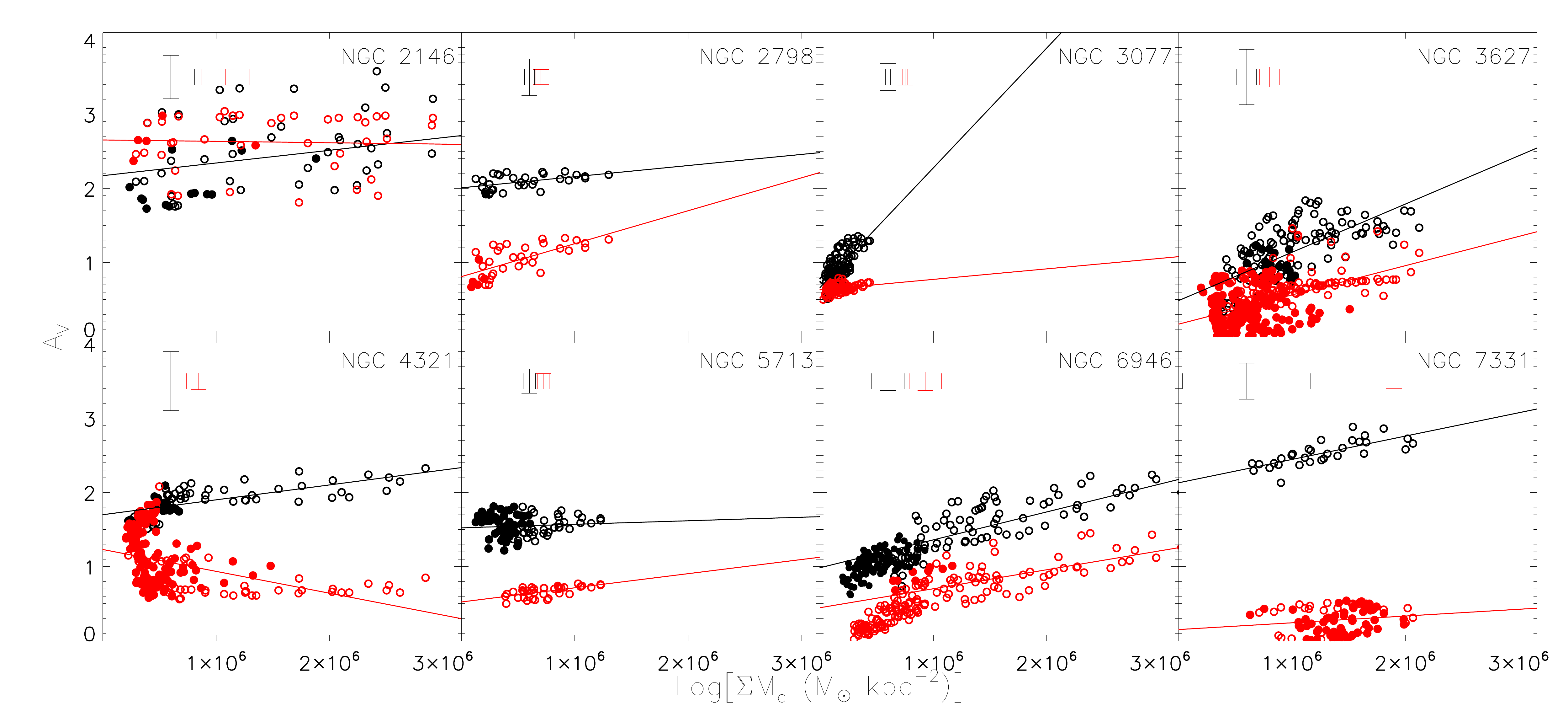}
\caption{Optical extinction derived from the Balmer decrement (black) and from the stellar continuum (red) as a function of the dust mass surface density at the convolved 18\arcsec\ resolution for 6\arcsec\ square pixels for each of the 8 galaxies.  Measurements that are available using only one of the two techniques, and thus provide complementary information, are shown with filled symbols.  All galaxies show a roughly linear relation. Lines show a bisector fit to guide the eye.  Median error bars for each are shown in the upper left corner.
\label{fig:pix_each}}
\end{figure*}

Figure \ref{fig:ngc3627} shows an example comparison of the dust mass surface density maps with the high and low resolution A$_V$ maps for NGC 3627.  
The pixel by pixel comparison is shown in Figure \ref{fig:pix_each} for each of the eight target galaxies using both the Balmer decrement reddening (black) and the stellar continuum reddening (red).  In some cases, particularly NGC 3627, NGC 4321 and NGC 7331, we observe no H$\alpha$ emission  in the central region where the stellar emission is otherwise bright.  In these regions, although we have no measure of the Balmer decrement, the stellar continuum reddening provides complementary information for these missing pixels (filled red).  We also find these two measures to be complementary in the outer disk, for example with the extranuclear pointing in NGC 6946, where bright H\textsc{ii} regions allow measurement of the Balmer decrement although the stellar continuum light is negligible (filled black).  In each galaxy we note that the regions with higher dust mass surface densities do tend to have higher visual extinction.
  We note a stronger trend for A$_{V,g}$ than A$_{V,s}$, suggesting that the attenuated emission lines are more sensitive overall tracers of the dust mass in these galaxies.    

We note here a few of the unusual features particular to the galaxies in our sample.  

\textit{NGC 7331},  as we mentioned before, has a strong difference between the near (west) and far (east) sides of the disk due to its high inclination, with the stars populating spiral arms and a bright bulge but most of the gas and dust located in a large-scale ring \citep{Regan2004}.  This can explain some of the large discrepancy between A$_{V,s}$ and A$_{V,g}$, in particular as the foreground bulge light appears to dominate in the stellar continuum compared to light from the far side of the disk where the dust resides.  This, however, does not entirely explain the large difference in A$_V$ values on the near side.  
The luminosity-weighted median stellar age from the SSP template fits to the star forming regions in NGC 7331 are slightly higher (4.5 Gyr) than in similar regions for other galaxies in our sample (3-4 Gyr).  This  suggests that the bulge light, which exhibits median luminosity-weighted stellar ages of $\sim$10 Gyr in the galaxy center, may have a non-negligible contribution to even the near side of the disk that results in stellar continuum light that is more heavily dominated by un-attenuated bulge stars.  This increase in attenuation within the bulge relative to the disk has also been reproduced in radiative transfer modeling of decomposed bulge-disk systems \citep{Tuffs2004}.

\textit{NGC 4321} exhibits particularly high dust mass surface densities in the very central region, however it is known to have an AGN which may affect the dust grain size distribution and resulting dust mass estimates. Here we have not adapted our Balmer decrement conversion to account for the AGN, however the effect of that correction on A$_{V,g}$ would be minor.   It also has a bright circumnuclear ring with a set of very dusty HII regions \citep{Kennicutt1989, Knapen1995}. 

\textit{NGC 2146} displays quite good agreement between A$_{V,s}$ and A$_{V,g}$. This LIRG galaxy is known to have strong outflowing winds \citep{Armus1995}, which may contribute to unusual mixing of the dust, foreground outflows, and a shocked gas contribution to the Balmer emission. 

\begin{figure*}[b!]
\centering
\includegraphics[width=3.5in]{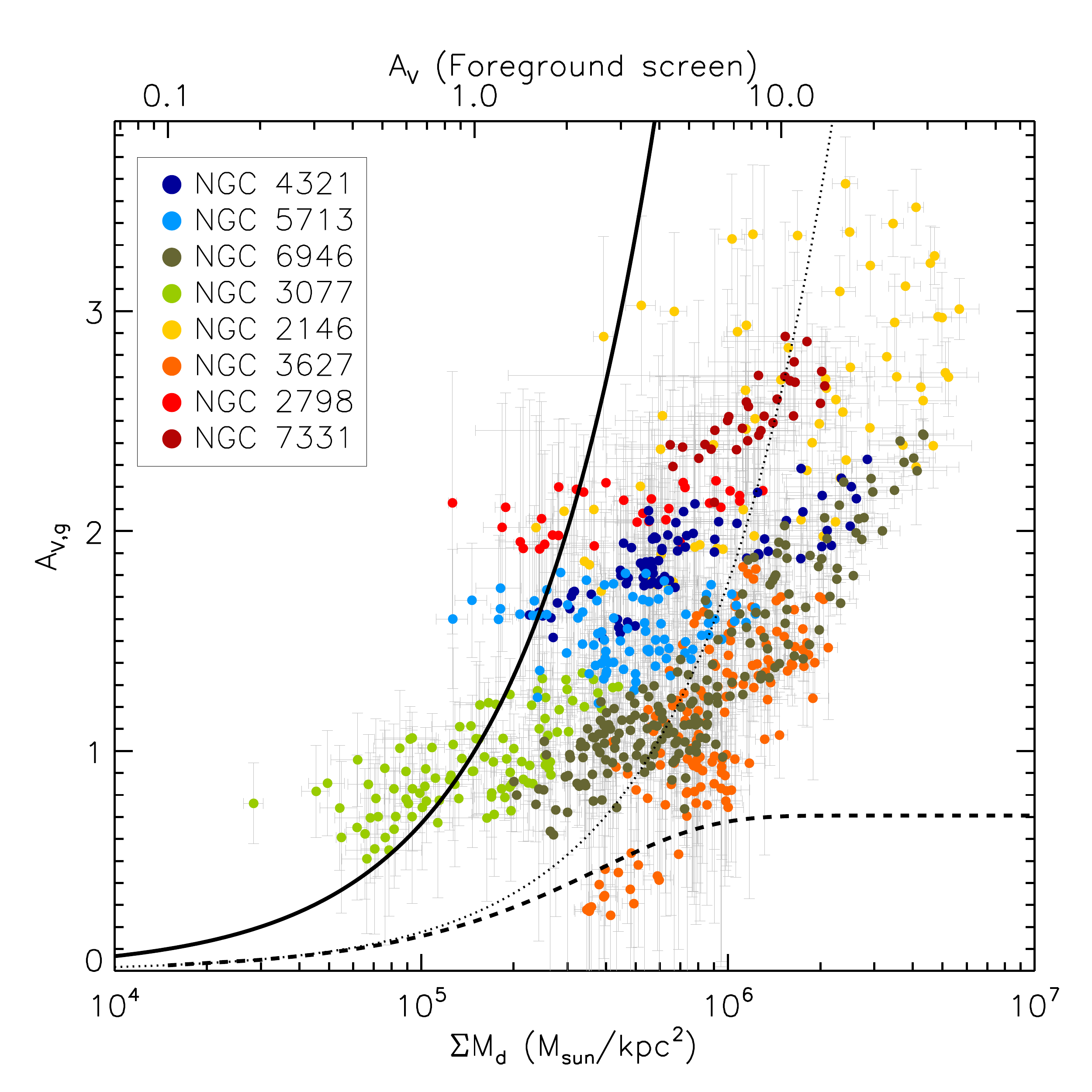}
\includegraphics[width=3.5in]{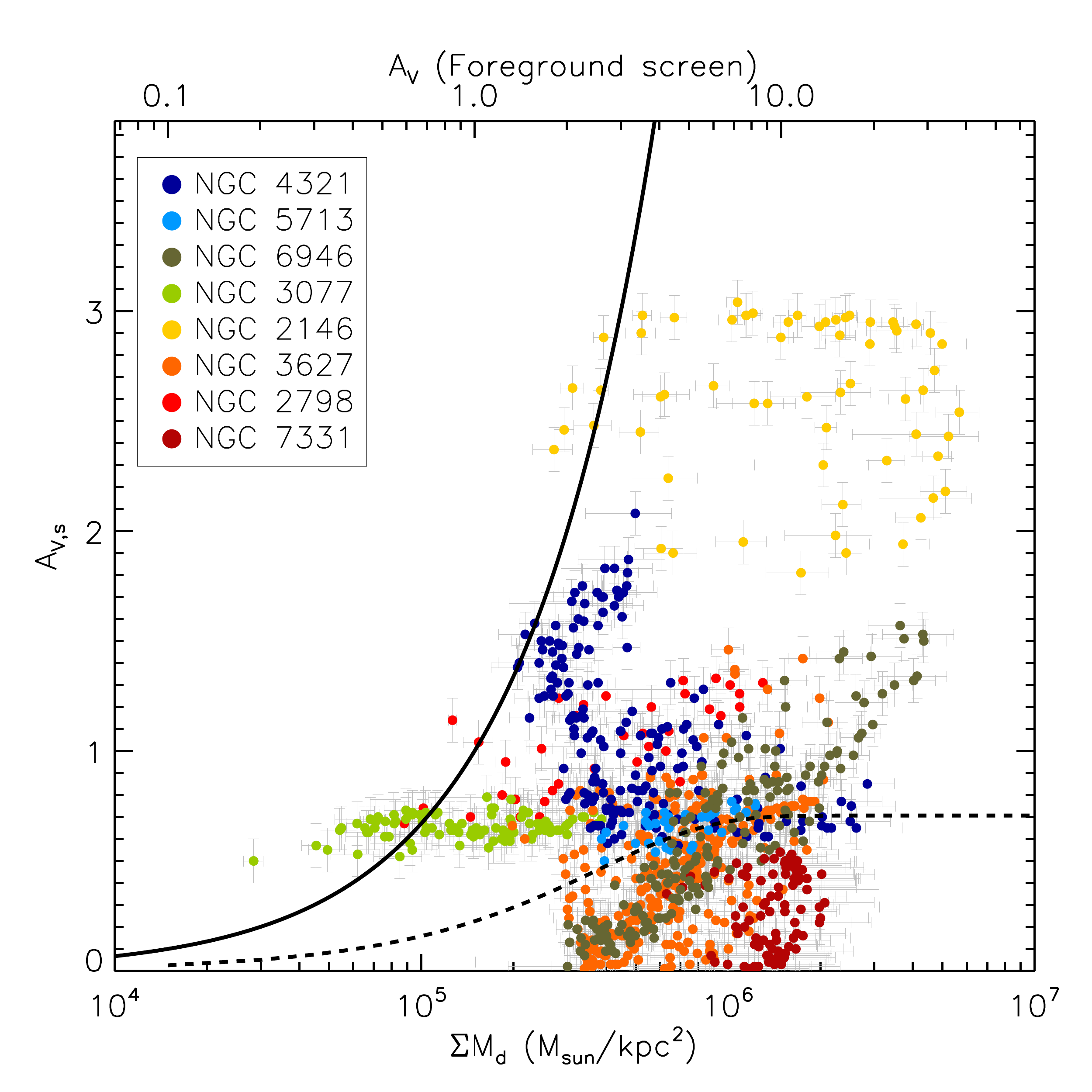}
\caption{ Comparison of the dust mass surface density with the V-band extinction.  On the left, we measure the visual extinction from the Balmer decrement, while on the right we use instead the reddening of the stellar continuum. The solid line is the dust mass surface density implied assuming a foreground screen model, the dashed line assumes a mixed media model.  Conversion of the dust mass surface density into a visual A$_V$ assuming all of the dust is in a foreground screen is shown on the top axis.  A best fit to the correlation observed from the Balmer decrement by scaling the foreground screen model (dotted line) by a factor of 3.8 is able to predict the dust mass surface density within a factor of two for $\sim$70\% of the sample.  Any correlation in the extinctions from the stellar continuum reddening is much less clear.  Galaxies are color coded roughly by inclination, where red-orange-yellow galaxies are more edge-on and blue-green galaxies are more face on, and are listed in order from most face-on to most edge-on in the legend.  
\label{fig:av_all}}       
\end{figure*}

In Figure \ref{fig:av_all} we show A$_{V,g}$ and A$_{V,s}$ from all eight galaxies together as a function of the dust mass surface density at that position.  We consider two simple models for the dust geometry that bracket the possible extremes.  A more realistic treatment of the complex geometry between stars and dust would require detailed radiative transfer models \citep{Witt2000, Tuffs2004}, such as have been applied to the edge-on galaxy NGC 891 \citep{Kylafis1987, Xilouris1998, Bianchi2008, Schechtman-Rook2012}, and we will pursue this approach in future work.

One simple model would be a configuration where all the dust falls in a uniform screen between the emitter and the absorber.  For this case, we convert between dust mass surface density and A$_V$ by assuming the observed Milky Way ratio of visual extinction to Hydrogen column, and a fixed dust to gas mass ratio.  Here we use A$_V$/N$_H$ = $5.34 \times 10^{-22}$ mag cm$^2$/H for a dust/H ratio $\Sigma M_d/(N_H m_H) = 0.010$ (see Table 3 of \citealt{Draine2007}), which gives
\begin{equation}
\label{eqn:screen}
A_{V,{\rm screen}} = 0.67 \frac{\Sigma M_{d}}{10^5 ~M_\sun ~kpc^{-2}} mag
\end{equation}
In practice, for a measured A$_V$ the corresponding dust mass surface density is expected to be a lower limit as the illuminating sources are most likely within the dust distribution.  Given that the highest dust mass surface densities observed in our sample would result in extinctions of up to 40 magnitudes, we also expect that realistically there must be a maximal optical depth to which we are sensitive.

At the opposite extreme, we can also consider a mixed media model where stars and dust are uniformly mixed with isotropic scattering \citep{Calzetti2000}.  In this case, the observed V-band extinction achieves an upper limit as an effect of scattering in the dust clouds and saturation that leads to a grey attenuation law.   In this situation the resulting effective extinction can be described via:
\begin{equation}
A_{V,mixed} = R_V \frac{1.086}{k(H\beta)-k(H\alpha)} ln\left[\frac{\gamma(H\alpha)}{\gamma(H\beta)}\right]
\end{equation}
\begin{equation}
\gamma(\lambda) = \frac{1-e^{-\tau_{sc}(\lambda)}}{\tau_{sc}(\lambda)}
\end{equation}
\begin{equation}
\tau_{sc}(\lambda) = \sqrt{1-\omega_\lambda} 0.921 A_{V,screen} / R_V
\end{equation}
where $\tau_{sc}$ models the optical depth given scattering within the dust, and $\gamma$ functionally limits the value of $\tau_{sc}$ in the high optical depth case.  Here we take $\omega_\lambda = -0.48 ~{\rm log}_{10}(\lambda/{\rm \AA})+2.41$ as the albedo adopted by \cite{Calzetti1994} for optical wavelengths 3460\AA~ to 7000\AA, R$_V$ as the typical Milky Way value of 3.1, $k(\lambda)$ as taken from the \cite{Calzetti2000} attenuation law, and we compute A$_{V,screen}$ following Equation \ref{eqn:screen}.

As seen on the left side of Figure \ref{fig:av_all}, the reddening of the Balmer line emission generally increases with increased dust mass surface density.  This is consistent with the results of \cite{Munoz-Mateos2009b}, who used the total IR to UV ratio to estimate the attenuation and Spitzer imaging to trace radial dust properties within nearby galaxies.  Their attenuation tracer is in between our two tracers, as the stars contributing to the bulk of the UV emission are not as young and dust-enshrouded as those responsible of the nebular emission, but not so evolved as those dominating the optical stellar continuum.   
Our points fall consistently within the bounds for the two models we consider.  This suggests that all recombination line emitting regions are affected by a combination of mixed dust and emitting material as well as a foreground screen component that allows for A$_{V,g}$ values higher than the mixed media model upper limit.  

Three galaxies, NGC 2798, NGC 5713 and NGC 3077, have regions that even given the estimated errors would appear to be forbidden by a foreground screen model, having significantly more optical attenuation than would be expected given the amount of dust detected in emission.   Closer examination of NGC 2798 reveals that these points fall on the eastern side of the galaxy, where the peak in optical attenuation appears shifted in the direction of a close interacting companion, NGC 2799.  It also corresponds to a slight reddening in the g-r color seen in SDSS images, suggesting that this may correspond to some overlap region between the two galaxies.  Given that this is the farthest galaxy and has relatively low (750 pc) spatial resolution, a sufficiently narrow dust feature encompasing the bright H$\alpha$ emission could result in significant reddening of the emitting region while minimally affecting the surface density of the dust.  NGC 5713 is the second most distant galaxy in our sample, and may suffer a similar effect.  NGC 3077, on the other hand, is the nearest galaxy in our sample, however it is also part of an interacting system and is well known for its unusual dust morphology \citep{Walter2011}. We investigate in more detail the importance of spatial resolution in Section \ref{sec:geometry}.

Often the foreground screen assumption is used to estimate the dust mass, with a factor of two or three higher dust masses assumed to account for dust on the far side of the galaxy and other geometrical effects.  We find that a slightly higher factor is required, approximately a factor of four, to bring the foreground screen model into better agreement with the observations.  We discuss this and provide an empirical relation in Section \ref{sec:empirical}.  

The reddening of the stellar continuum shows no such clear correlation with the dust mass surface density (Figure \ref{fig:av_all}, right),  suggesting that the stars, though more completely sampling the galaxy, are not well sampling the dust distribution. This is additionally seen in Figure \ref{fig:pix_each}, as in most galaxies the stellar reddening is a less sensitive tracer of the dust than the Balmer line reddening. 
\\
\\
\section{Discussion} 
\label{sec:discussion}
The flexibility of our IFS data allow us to discuss spectroscopic properties in our sample on two scales - the full PPAK resolution of $\sim$2\farcs5 (physical scales of 20-100 pc) and the convolved lower resolution 18\arcsec\ SPIRE 250 scale (350 pc - 2 kpc). Here we take both into account and address the correlations observed in the lower resolution comparison with dust as seen in emission, particularly given our added knowledge of the high resolution distribution of the dust in absorption.

\subsection{Possible drivers of scatter within and between galaxies} 
\label{sec:geometry}
Considered individually (Figure \ref{fig:pix_each}), all galaxies show a correlation between higher dust mass and higher A$_{V, g}$, and most show a similar correlation with A$_{V,s}$.  Considered together (Figure \ref{fig:av_all}), a more universal correlation between the Balmer line reddening and the dust mass surface density is seen, however no such relation is clear in the stellar continuum reddening. This suggests that the stellar continuum is not well sampling the dust distribution.  This may be attributable to the preferential location of the dust in star-forming giant molecular clouds or to the thinness of the dust layer, both of which may cause the more extended distribution of stars within the galaxy to suffer irregular reddening effects from the dust.  

On the other hand, all galaxies show a similar correlation in A$_{V,g}$. In our two mosaicked galaxies, NGC 4321 and NGC 3627, we probe nuclear, bar and spiral arm environments in a spatially resolved way and see no stark discontinuities between different regions. NGC 6946, for which we probe both the central and an extra nuclear region at $\sim$0.4 R$_{25}$, also displays a fairly uniform relation throughout the galaxy.

\begin{figure}[t!]
\centering
\includegraphics[width=3.5in]{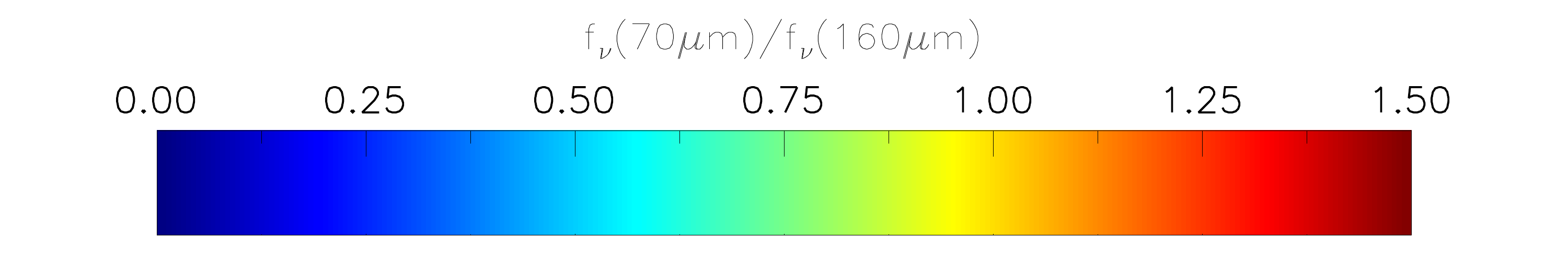}
\includegraphics[width=3in]{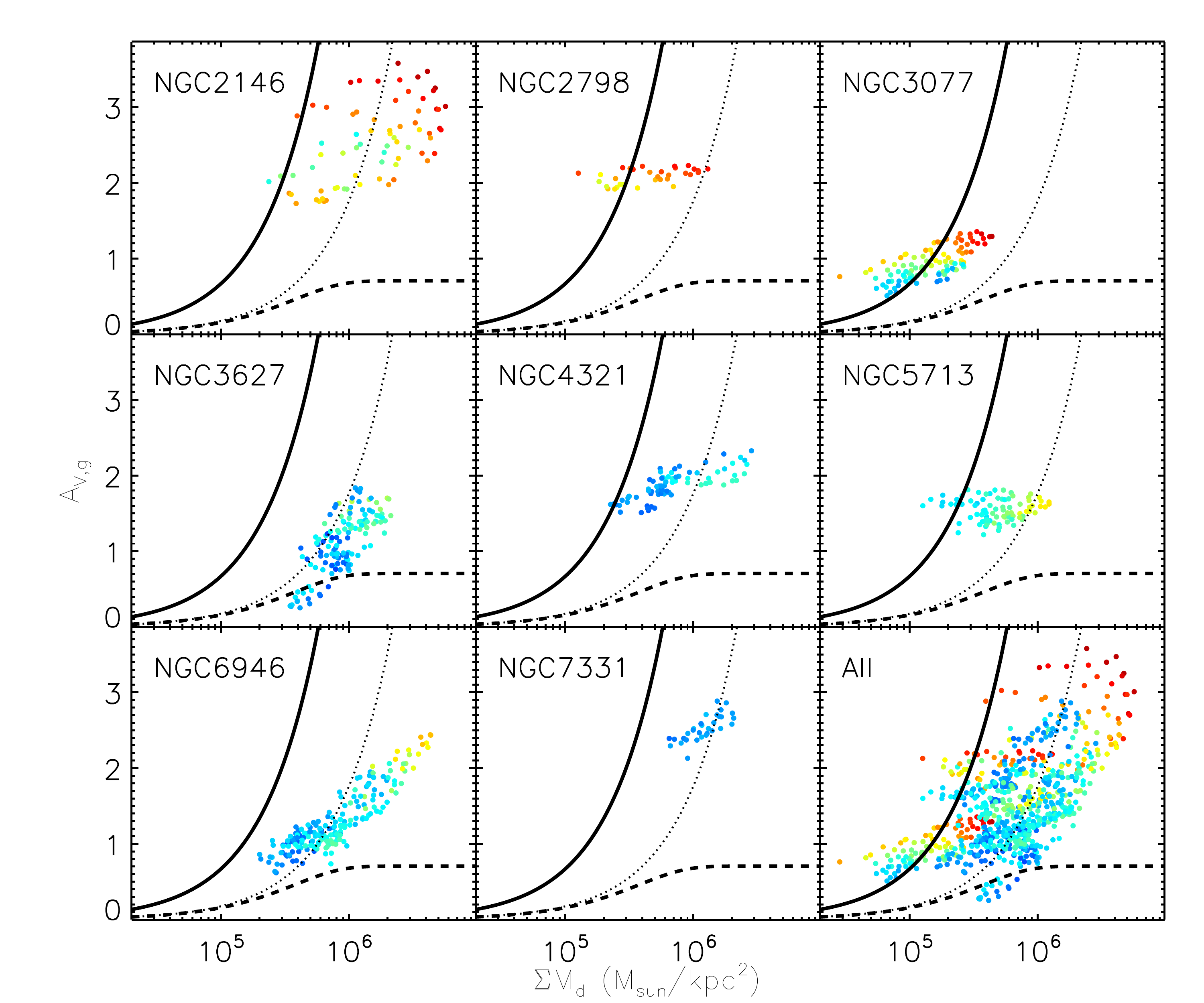}
\caption{Trends with the 70 $\mu$m to 160 $\mu$m ratio for each of the galaxies considered individually and together (bottom right).   Correlation within galaxies appear to be driven by the well established connections between dust, gas, star formation and heating, however this and other similar tracers of SFR and dust heating cannot explain the scatter observed between galaxies.
\label{fig:trends}}
\end{figure}

Within each galaxy, much of the correlation appears to be driven by the well established connections between dust, gas, star formation and heating.  Figure \ref{fig:trends} shows one example of this.  We color code each region by the ratio of PACS 70$\mu$m to PACS 160$\mu$m emission, which traces the dust heating \citep{Dale2005, Dale2007}, and find clear correlations within each galaxy, with the dustier and more extincted regions containing relatively warmer dust. These results are robust to our choice of dust heating tracer, including parameters fit through the dust modeling (i.e. $\gamma$, the fraction of dust heated over the background level interstellar radiation field, or $f_{PDR}$, the fraction of dust heated by PDRs, \citealt{Aniano2012}) or tracers of the DIG component (i.e. [SII]$\lambda$6716/H$\alpha$, see Figure \ref{fig:sfrsd}).  
The relation between higher SRF surface density (or H\textsc{ii} region fraction) and higher A$_{V,g}$ is most likely connected by a higher gas surface density, leading to greater dust columns (A$_V$) and a higher SRF via the Kennicutt-Schmidt relation.  
This is further supported by work directly comparing the attenuation with the gas surface density \citep{Boquien2013}, which shows a similar general correlation between A$_V$ and gas surface density but with increased scatter.  

While these correlations with dust heating can explain some of the trends observed within galaxies, it cannot account for the differences observed between galaxies (Figure \ref{fig:trends}, bottom right), and our sample of galaxies, taken together, displays significant scatter. As our sample consists mainly of L$_*$ spiral galaxies, we can exclude effects from differences in metallicity or stellar mass.  For example, all of our galaxies have circular velocities well above the 120 km\,s$^{-1}$ threshold identified by \cite{Dalcanton2004} as tracing structural changes in the ISM and dust distribution.  Variations by Hubble type have been shown to indicate differences in the dust distribution \citep{Dale2007, Munoz-Mateos2009b}, with later type spirals exhibiting clumpier dust morphologies and less attenuation, however these can be difficult to disentangle from the effect of metallicity which also correlates with Hubble type \citep{Moustakas2010}.   We are here limited by our small sample, which contains only one late-type spiral (NGC 6946 with type SABcd).  We do note that this galaxy exhibits relatively lower attenuation at fixed dust mass surface density, similar to the trend seen in Figure 10 of \cite{Munoz-Mateos2009b}, however given it is included within the scatter between the remaining early-type spiral galaxies we believe Hubble type cannot be the main driver of scatter in our sample.

Another property to consider is the effect of inclination.  Indeed, some of the more inclined galaxies do achieve high extinctions, however the highest values (A$_{V,g}\sim$3 for NGC 2146) do not come from the most inclined galaxy (NGC 7331).  We also note that for the inclined galaxies, although each line of sight projects through a range of disk environments they still exhibit relatively tight correlations. The most face-on galaxies in the sample (NGC 6946, NGC 5713 and NGC 4321) also do not form a particularly uniform sequence when taken together. 

There is also no clear connection between inclination and dust mass surface density, although we have made no attempt to correct the measured surface densities to face-on values.  NGC 4321 and NGC 6946, both fairly face-on, achieve some of the highest dust mass surface densities, similar to the dusty starburst galaxy NGC 2146.  In NGC 4321 we note that these pixels fall directly in the galaxy center, where there is both an AGN source, which could be altering the dust grain composition and dust temperature and affecting our dust mass estimate, as well as a nuclear ring, which may be particularly dust rich. The A$_{V,g}$ and $\Sigma$M$_d$ measured for this region of this galaxy is not an outlier when compared to the rest of the sample. NGC 6946 has no known central AGN, but other galaxies which do (NGC 3627 and NGC 7331) show no central peak in the dust mass surface densities.  

A slight effect from the galaxy geometry may be seen between the near and far sides of NGC 3627 and NGC 4321, where the H\textsc{ii} regions in the south-east arm and north-west arm (respectively) at the connection with the bar (Figure \ref{fig:optical}) exhibit the same extinction in the stellar continuum as in the Balmer decrement. NGC 4321 unfortunately has the stellar continuum for only one side, however NGC 3627 is somewhat clearer.  This suggests that the arm with equal extinction in A$_{V,g}$ and A$_{V,s}$ is on the near side, as the stellar light must come mostly from the background disk and probes a similar dust column as the Balmer lines. 

\begin{figure*}[t!]
\centering
\includegraphics[width=6in]{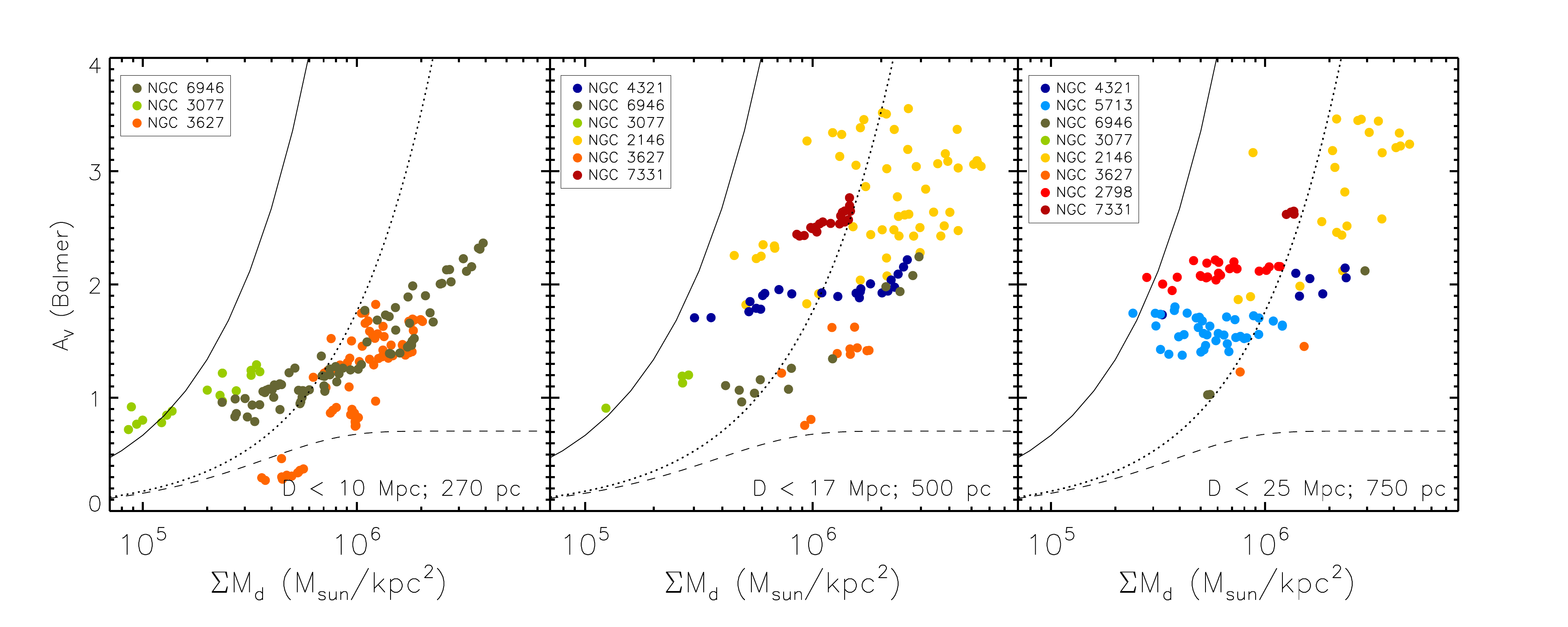}
\caption{ Comparison of the dust mass surface density with the visual extinction where apertures in each galaxy are scaled to the same physical size.  The limiting distance and corresponding physical scale for each is shown in the bottom left. Lines indicate limiting models of dust geometry, as described in \ref{fig:av_all}.  Galaxies are color coded and listed in order of increasing inclination angle.       
\label{fig:physcl}}
\end{figure*}

A more important factor may be the physical size scales probed in each galaxy. The A$_{V,g}$ in the farthest galaxies has little or no dependence on dust mass surface density, most apparent in Figure \ref{fig:pix_each} for the most distant galaxies NGC 2798 and NCC 5713.    We further test this by sampling the line maps for each galaxy with apertures of equal physical size (Figure \ref{fig:physcl}). The furthest galaxy in the sample, NGC 2798 at 25.8 Mpc, gives a physical scale of $\sim$750 pc.  Probing an equivalent angular size is prohibitive for many of the galaxies in our sample given the 1\arcmin~ IFS field of view, and many of our targets are reduced to a single point.  Consequently, much of the correlation between dust in emission and absorption (Figure \ref{fig:physcl}, right) is removed, particularly if we exclude the unusual and dust rich NGC 2146, and as such we caution against using global corrections or scales larger than 1 kpc to measure dust mass from a single A$_V$ measure. Moving to smaller physical scales, the $\Sigma$M$_d$ and A$_{V,g}$ do begin to trace each other with significantly reduced scatter for the three galaxies we can observe at 270 pc scales. This is consistent with the scale height of dust observed in spiral galaxies \citep{Xilouris1999}, and suggests that non-uniformity in the dust distribution on these scales contributes to the scatter in the relation we derive for the range of scales.  

The variations in the relation between the reddening of the stellar continuum and the reddening of the emission lines that depend on the H$\alpha$ flux are also resolved out at these scales (see also Section \ref{sec:absorption}), however two galaxies stand out.  NGC 7331, a highly inclined galaxy, and NGC 2146, a dusty starburst galaxy, both exhibit strong dust lane features, however they exhibit very different stellar reddening compared to the emission lines.  NGC 7331 appears to have stellar light dominated by a bright bulge component, seen also in the mean stellar age as determined from spectral fitting.  This results in the stellar light tracing almost none of the dust that exists in a thin star-forming disk that in our small field of view is not well separated from the bulge component, producing exaggerated near- and far- side effects due to the large inclination angle. NGC 2146 has a strong starburst and there is evidence of winds and outflows, possibly resulting in mixing of the dust or significant ejection of dust that now sits foreground to both the stellar and emission line sources.  This is the opposite effect as would be expected given the moderately high (i$\sim$60$^\circ$) inclination.  These effects are also seen for these two galaxies in the high resolution maps (Figure \ref{fig:sfrsd}).

Despite these complications from geometric effects, we note that the simple limiting models from \cite{Calzetti2000} of dust in a foreground screen or in a mixed media model well bracket our results for the Balmer line reddening.  This suggests that some combination of these two effects, which may vary from galaxy to galaxy,  can largely explain our results without recourse to more detailed radiative transfer models, which is not the case for the stellar continuum reddening. 

\subsection{Stellar continuum versus emission lines as reddening tracers} 
\label{sec:absorption}
\begin{figure*}
\includegraphics[width=7in]{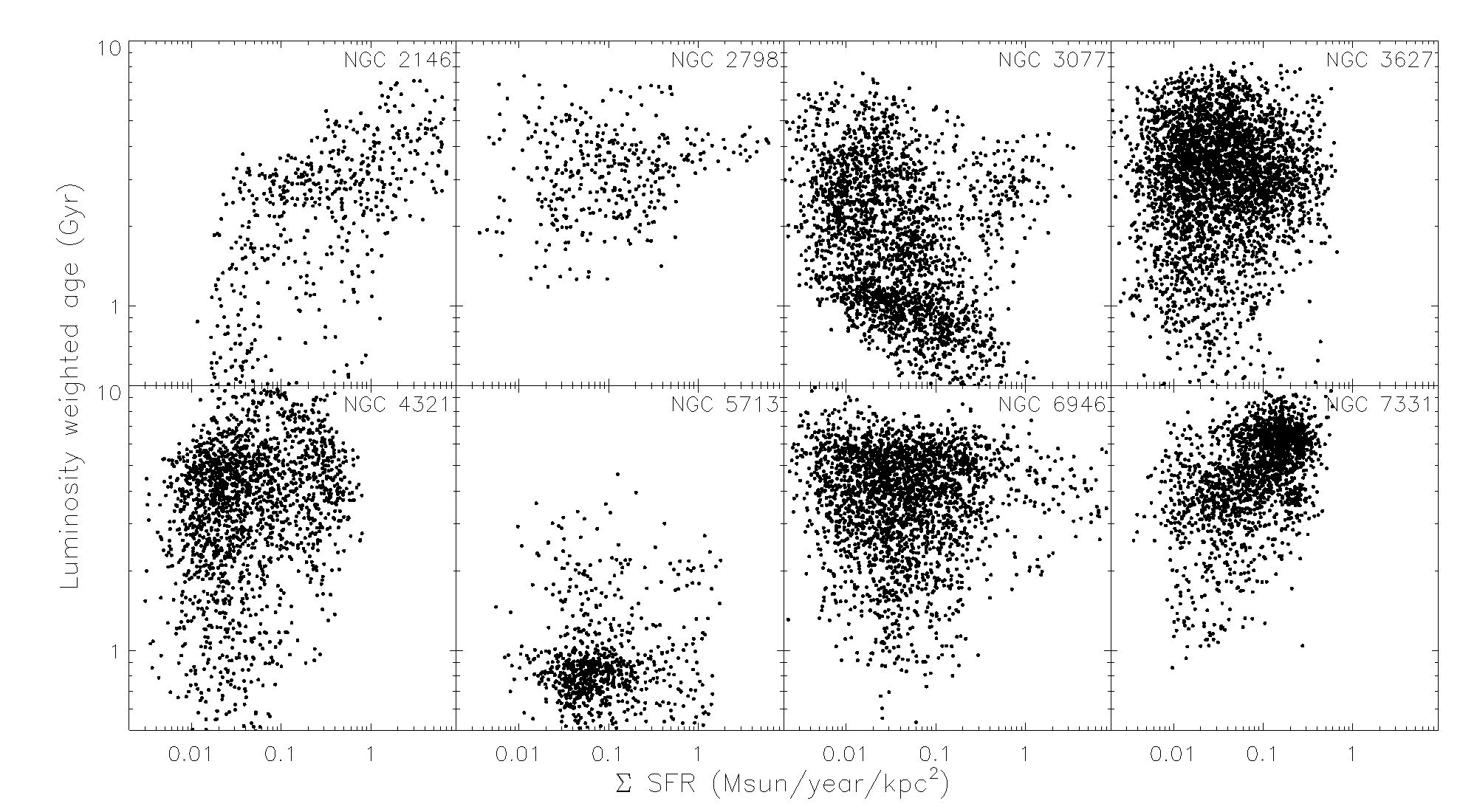}
\caption{ The luminosity-weighted stellar age as a function of the star formation rate surface density for each position in our galaxy disks.  Aside from NGC 3077, which is unusual for being the most nearby dwarf galaxy in our sample, we observe no correlation, suggesting that the bright H\textsc{ii} regions are not dominating the stellar continuum light in their vicinity and the stellar continuum and Balmer line emission provide independent spatial sampling of the dust distribution. 
\label{fig:sfrage}}
\end{figure*}

In our high resolution A$_{V,s}$ and A$_{V,g}$  maps we see a clear difference of about a factor of two between the reddening affecting the Balmer emission lines and the reddening affecting the stellar light.  This has been reported previously for individual H\textsc{ii} regions \citep{Calzetti1994} and global observations of star-forming galaxies \citep{Hao2011}, and is generally attributed to the preferential location of H$\alpha$ emission within dusty clouds from which the young stars providing the ionizing photons have formed \citep{Calzetti2000,Charlot2000}. As a result, we find that the Balmer emission line reddening traces more sensitively the total dust mass surface density compared to the stellar continuum reddening (see Figure \ref{fig:av_all}). 
We check that the stars and the H$\alpha$ bright regions independently sample the dust distribution through the luminosity-weighted stellar ages.  At each position, our SSP template fits provide some constraint on the type of stellar population dominating the light in that region.  We find no systematic correlation between the SFR surface density and the luminosity-weighted age (Figure \ref{fig:sfrage}), suggesting that within our galaxies the bright H\textsc{ii} regions are not typically dominating the stellar continuum light in their vicinity and thus we may consider the Balmer line emission and the stellar continuum as completely independent probes.  NGC 3077 does show some evidence for lower stellar ages in regions of high star formation, with large scatter.  It is, however, the galaxy for which we have the highest spatial resolution, as well as amongst the least massive in our sample, both of which may factor into this difference.

We further explore the environmental difference of these two probes through the low-ionization [S\textsc{ii}]$\lambda$6716/H$\alpha$ line ratio, which is sensitive to the transition between H\textsc{ii} and diffuse ionized gas (DIG) regions \citep{Reynolds1985,Haffner1999,Rand1997,Hoopes2003}.  
Observations of these lines towards H$\alpha$ emitting regions in the Milky Way suggest a transition at approximately [S\textsc{ii}]$\lambda$6716/H$\alpha = 0.2$, with H\textsc{ii} regions typically exhibiting the lower line ratios \citep{Madsen2006}.  
The lower panels of Figure \ref{fig:sfrsd} shows the ratio A$_{V,s}$/A$_{V,g}$ as a function of the extinction corrected H$\alpha$ flux, color coded by the [S\textsc{ii}]$\lambda$6716/H$\alpha$ ratio, and a clear correlation of lower line ratios to higher H$\alpha$ fluxes is apparent in each galaxy.  This has also been seen in other nearby galaxies \citep{Wang1998, Blanc2009}.
Separating our sample at  [S\textsc{ii}]$\lambda$6716/H$\alpha$ = 0.2, we do find a lower  A$_{V,s}$/A$_{V,g}$ ratio of 0.5 for H\textsc{ii} dominated regions compared to 0.7 for the DIG dominated regions.  This result is not sensitive to our specific choice of [S\textsc{ii}]$\lambda$6716/H$\alpha$ ratio, and suggests that it is H$\alpha$ emission originating from H\textsc{ii} regions that is preferentially dustier, while diffuse H$\alpha$, like the stellar distribution, is more uniformly mixed.
We further note that the transition in the ratio A$_{V,s}$/A$_{V,g}$ occurs fairly systematically at a fixed $\Sigma_{SFR}$ of approximately 0.1 M$_\sun$ yr$^{-1}$ kpc$^{-2}$.  This also largely corresponds to the division between H\textsc{ii} and DIG regions as determined through the [S\textsc{ii}]$\lambda$6716/H$\alpha$ ratio, though for NGC 2146 and NGC 2798 this occurs at slightly higher $\Sigma_{SFR}$.  

Given the lack of high $\Sigma_{SFR}$ regions having low A$_{V,g}$ (Figure \ref{fig:sfrsd}), we believe this is strong evidence that H\textsc{ii} regions must be well correlated with the dust on 20-120 pc scales.  We note that on these small scales we cannot speak to the presence of dust that exist without H\textsc{ii} regions \citep[see]{Bergin2007}, however we find systematically that the brightest H\textsc{ii} regions always experience attenuation (see Figure \ref{fig:sfrsd}).  
In this work we trace only the ionized gas and dust, however our conclusion is very different than has been drawn for correlations of molecular gas with H\textsc{ii} regions.  Recent high resolution observations of giant molecular clouds in nearby galaxies suggest that the star formation law that robustly relates $\Sigma_{SFR}$ and gas surface density for $\sim$1 kpc spatial scales breaks down on $\sim$80-300 pc scales \citep{Schruba2010, Onodera2010}.  This change is attributed to a shift from observing a large number of H\textsc{ii} regions together to observing individual regions, which will have varying properties depending on the state of evolution.  We will address this issue more fully in future work that compares the CO gas distribution with our dust extinction maps.

\subsection{Empirical relation} 
\label{sec:empirical}
Given that in most situations some measurement of the optical extinction is available and we wish to constrain the dust mass, we determine an empirical relation between the dust in absorption and dust in emission for our full sample of eight galaxies.  Typical geometric corrections to a foreground screen model used in the literature assume 2-3 times more dust, which accounts for midplane sources.  We find that scaling the foreground screen relation (Equation \ref{eqn:screen}) by a factor of 3.8 gives the best fit (dotted line, Figures \ref{fig:av_all} and \ref{fig:physcl}), and is able to predict the dust mass surface density given A$_{V,g}$ for 67\% of the sample to within a factor of two, and 90\% of the sample to within a factor of three.  Direct fits to the data neglecting physical motivations, done both in linear and log space for the dust mass surface densities, do not result in significantly better fits and so we recommend the following conversion from A$_{V,g}$ to dust mass surface density

\begin{equation}
\Sigma M_d = 5.7 \times 10^5 A_{V,g}~M_\sun~kpc^{-2}
\end{equation}

at physical scales of 350 pc - 2 kpc.

Taking our conclusions about geometrical effects into account, we believe that much of the scatter may be introduced by non-uniformity of the dust on physical scales smaller than 500 pc.  Additional scatter at higher extinctions comes from inclination effects, which may allow for an increased foreground screen contribution.  We also have a few spurious points which fall above the expected bounds for the foreground screen model, a scatter which is largely accounted for by the errors.

When comparing stellar continuum extinction with line extinction, a factor of $0.44 \pm 0.03$ \citep{Calzetti2000} is typically used to correct for the embedding of these emission line sources, as measured from starburst galaxies and bright H\textsc{ii} regions.  For our sample, we have the ability to refine our selection of regions and galaxies to optimally exclude complications from galaxy geometry and H$\beta$ detection limits.  Considering only those H$\alpha$ bright H\textsc{ii} regions with $\Sigma_{SFR}$ above 0.1 M$_\sun$ yr$^{-1}$ kpc$^{-2}$, where the scatter in the A$_V$ ratio is significantly reduced,  and including only the more face-on galaxies in our sample (NGC 3077, NGC 4321, NGC 5713, NGC 6946) we find  
\begin{equation}
A_{V,s} = ( 0.47 \pm 0.006 ) A_{V,g}
\end{equation}
at physical scales of 20-100 pc.

As bright H\textsc{ii} regions make up 40-70\% of the H$\alpha$ flux in galaxies \citep{Oey2007, Kennicutt2012}, this relation would be an appropriate choice for integrated spectra of star-forming galaxies.  In the case of resolved regions, some care should be taken to ensure an appropriate dust tracer is used. H$\alpha$ fluxes should be corrected using line emission tracers of the reddening, or appropriately corrected stellar reddening tracers.  We caution that a ratio of 0.7 may be more suitable for DIG dominated regions and galaxies.  Alternately, for dust correction to continuum light, we recommend that corrections to extinctions derived from Balmer line emission also be taken into account.
\\
\section{Conclusion} 
\label{sec:conclusion}
We examine the distribution of the dust in emission, using Herschel imaging as part of the KINGFISH project and fits of \cite{Draine2007} dust models, and compare this to the distribution of optical reddening, using IFS imaging by the PMAS instrument in PPAK mode, of eight nearby galaxies.  We find a correlation between the Balmer line reddening and the dust mass surface density on physically resolved (300-700 pc) scales in these galaxies, which corresponds to a factor of 3.8 larger dust mass than would be expected given the observed Balmer decrement  for a foreground screen of dust. 
We identify trends within galaxies connecting regions with increased dust mass surface density and attenuation to hot, star forming regions, as expect through the established correlations between dust, gas and star formation.    
We attribute most of the scatter that is observed between galaxies to the effects of differing physical scales due to different target distances, however we also note an inclination effect as the more inclined systems do exhibit the highest extinction.  We find no correlation between the stellar continuum reddening and the dust mass surface density, suggesting the stars are not well sampling the dust distribution.  Regions considered span a range of galaxy environments, including spiral arms, bars, nuclear and extra-nuclear regions, yet show relatively uniform relations within each galaxy.  Within the measurement uncertainties, all regions fall between two extremes in dust geometry, of either a completely foreground screen or a mixed media model, suggesting a combination of both exists in most galaxies. 

We also compare the effects of the extinction of stellar continuum light (A$_{V,s}$) with the Balmer decrement (A$_{V,g}$) at the finest physical scales (20-120 pc) allowed by our optical IFS data.  We reproduce the trend for higher A$_{V,g}$ by a factor of two, but note a transition at the lower SFR regions that are more diffuse gas dominated where we observe a ratio closer to unity. We attribute this difference in high H$\alpha$ emitting regions to the preferential location of H\textsc{ii} regions within dustier environments.  Two outliers from this relation, NGC 7331 and NGC 2798, represent two extremes of galaxy geometry.  NGC 7331 has a particular large and bright bulge, which appears to dominate the stellar continuum to produce very strong near- and far-side effects for which the Balmer emission is immune.  NGC 2146 shows fairly good agreement between the two A$_V$ measures, suggesting either better mixing of the dust with the stars or a completely foreground preponderance of dust, both of which may be caused by the strong starburst, winds and shocks observed in this galaxy.  
 
While the ability of dust in absorption to trace dust in emission is relatively reliable on these physical scales, we caution against use of integrated measures to infer global dust properties. We find that even within the range of distances and scales probed by our sample the sensitivity of A$_V$ to dust mass changes, as in our most distant galaxies NGC 2146 and NGC 5713 the measured A$_{V,g}$ is relatively insensitive to the dust mass surface density. We note that at physical scales of $\sim$750 pc the scatter in the relation is much larger than at $\sim$250 pc scales, which is consistent with the physical scales of dust lane features seen in optical images. However, given the general predictive ability of the optical A$_{V,g}$ on all scales, measuring the dust mass within a factor of three for the majority of the sample, and with the advent of more IFS instruments, we believe that there is great potential to probe in more depth the dust properties of galaxies within the nearby universe with this method.

\acknowledgements
We would like to thank the referee for their helpful comments.  KK acknowledges the support of grants GR 3948/1-1 and SCHI 536/8-1 from the DFG Priority Program 1573, “The Physics of the Interstellar Medium”.

Based on observations collected at the Centro Astron\'{o}mico Hispano Alem\'{a}n (CAHA), operated jointly by the Max-Planck Institut f\"{u}r Astronomie and the Instituto de Astrofisica de Andalucia (CSIC).
Also based on observations made with Herschel.  Herschel is an ESA space observatory with science instruments provided by European-led Principal Investigator consortia and with important participation from NASA.
PACS has been developed by a consortium of institutes led by MPE (Germany) and including UVIE (Austria); KU Leuven, CSL, IMEC (Belgium); CEA, LAM (France); MPIA (Germany); INAF-IFSI/OAA/OAP/OAT, LENS, SISSA (Italy); IAC (Spain). This development has been supported by the funding agencies BMVIT (Austria), ESA-PRODEX (Belgium), CEA/CNES (France), DLR (Germany), ASI/INAF (Italy), and CICYT/MCYT (Spain).
SPIRE has been developed by a consortium of institutes led by Cardiff University (UK) and including Univ. Lethbridge (Canada); NAOC (China); CEA, LAM (France); IFSI, Univ. Padua (Italy); IAC (Spain); Stockholm Observatory (Sweden); Imperial College London, RAL, UCL-MSSL, UKATC, Univ. Sussex (UK); and Caltech, JPL, NHSC, Univ. Colorado (USA). This development has been supported by national funding agencies: CSA (Canada); NAOC (China); CEA, CNES, CNRS (France); ASI (Italy); MCINN (Spain); SNSB (Sweden); STFC (UK); and NASA (USA).

Funding for SDSS-III has been provided by the Alfred P. Sloan Foundation, the Participating Institutions, the National Science Foundation, and the U.S. Department of Energy Office of Science. The SDSS-III web site is http://www.sdss3.org/.

SDSS-III is managed by the Astrophysical Research Consortium for the Participating Institutions of the SDSS-III Collaboration including the University of Arizona, the Brazilian Participation Group, Brookhaven National Laboratory, University of Cambridge, Carnegie Mellon University, University of Florida, the French Participation Group, the German Participation Group, Harvard University, the Instituto de Astrofisica de Canarias, the Michigan State/Notre Dame/JINA Participation Group, Johns Hopkins University, Lawrence Berkeley National Laboratory, Max Planck Institute for Astrophysics, Max Planck Institute for Extraterrestrial Physics, New Mexico State University, New York University, Ohio State University, Pennsylvania State University, University of Portsmouth, Princeton University, the Spanish Participation Group, University of Tokyo, University of Utah, Vanderbilt University, University of Virginia, University of Washington, and Yale University.

This research has made use of the NASA/IPAC Extragalactic Database (NED) which is operated by the Jet Propulsion Laboratory, California Institute of Technology, under contract with the National Aeronautics and Space Administration.

\end{document}